\documentclass[Crown,sagev,times]{sagej}

\usepackage{moreverb,url}

\usepackage[colorlinks,bookmarksopen,bookmarksnumbered,citecolor=red,urlcolor=red]{hyperref}

\usepackage{bbold}
\usepackage{pgfplots}

\pgfplotsset{compat=1.18}
\usepgfplotslibrary{groupplots}

\usepackage{subcaption}
\usepackage{multirow}
\usepackage{graphicx}
\usepackage{booktabs}
\usepackage{mathtools }
\mathtoolsset{showonlyrefs=true}
\usepackage[section]{placeins}

\usepackage{xcolor}
\definecolor{vir1}{RGB}{253,231,37}   
\definecolor{vir2}{RGB}{94,201,98}    
\definecolor{vir3}{RGB}{33,145,140}   
\definecolor{vir4}{RGB}{59,82,139}    
\definecolor{vir5}{RGB}{68,1,84}      

\newcommand\BibTeX{{\rmfamily B\kern-.05em \textsc{i\kern-.025em b}\kern-.08em
T\kern-.1667em\lower.7ex\hbox{E}\kern-.125emX}}

\begin{document}

\runninghead{B. Farah et al.}

\title{\color{black}Designing clinical trials for the comparison of single and multiple quantiles with right-censored data\color{black}}

\author{Beatriz Farah\affilnum{1,2,3},Olivier Bouaziz\affilnum{2,4}\textsuperscript{*} and Aurélien Latouche\affilnum{1,5}\textsuperscript{*}}

\affiliation{%
\affilnum{1}INSERM U1331, Institut Curie, Saint-Cloud, France\\
\affilnum{2}CNRS, MAP5, Université Paris Cité, Paris, France\\
\affilnum{3}Université Paris-Saclay, UVSQ, Versailles, France\\
\affilnum{4}CNRS, UMR 8524 - Laboratoire Paul Painlevé, Univ. Lille, Lille, France\\
\affilnum{5}Conservatoire National des Arts et Métiers, Paris, France%
}

\corrauth{Beatriz Farah, INSERM U1331, Institut Curie, Mines Paris Tech, Paris/Saint-Cloud, France.\\ *These authors contributed equally to this work.}

\email{beatriz.farah@math.cnrs.fr}

\begin{abstract}
    \color{black}
    Based on the test for equality of quantiles originally introduced by Kosorok (1999), we propose new power formulas for the comparison of one quantile between two treatment groups, as well as for the comparison of a collection of quantiles.
    Under the null hypothesis of equality of quantiles, the test statistic follows asymptotically a normal distribution in the univariate case and a $\chi^2$ with $J$ degrees of freedom in the multivariate case, with $J$ the number of quantiles compared.
    The variance of the test statistic depends on the estimation of the probability density function of the distribution of failure times at the quantile being tested. 
    In order to apply the test on real data, we propose to estimate this quantity using a resampling-based method, as an alternative to Kosorok's original kernel density estimator.
    The whole procedure provides a practical tool for designing and analyzing data arising from clinical trials using quantiles of survival as an endpoint. 
    Simulation studies are performed to show the appropriateness of the power formulas. 
    We illustrate the proposed test in a phase III randomized clinical trial where the proportional hazards assumption between treatment arms does not hold.


    \color{black}


\end{abstract}


\maketitle

\section{Introduction}
In clinical studies with right-censored data, investigators have been increasingly interested in estimating the quantiles of the survival times, which are defined as being the smallest time when survival exceeds a threshold of interest~\cite{mboup2021insights, peng2021quantile}. As this measure is expressed in the timescale, the quantification of the benefit of one treatment arm over the other can be easily communicated and understood by clinicians and patients, in contrast to relative risk measures such as the hazard ratio which can be easily misinterpreted~\cite{hernan2010hazards, uno2015alternatives}.
In addition to allowing the benefit, if any, of the new treatment to be expressed in terms of time gained compared to the standard of care, the use of quantiles allows for robustness against outliers and does not depend on the shape of the survival distribution or the proportionality of the treatment effect.  
In particular, in immuno-oncology trials the usual assumption of proportional hazards between treatment arms is often not verified due to the delayed effect of immunotherapy and late separation of survival curves~\cite{mboup2021insights}. 
Given that the quantification of treatment effects using differences in quantiles inherently accounts for nonproportional hazards, this approach is particularly well-suited for such scenarios.

Few methods have been developed to compare quantiles of the survival function in the presence of censored data. Brookmeyer and Crowley (BC)~\cite{brookmeyer1982k} proposed a pooled-weighted Kaplan-Meier estimation approach in order to detect differences in the median survival times among several treatments. 
In order to avoid the estimation of the densities of the time-to-event in each treatment group, the authors proposed a simplified test statistic, which is only valid under the assumption that the survival distributions are equal in each group under the null hypothesis.
However, testing for homogeneity of survival distributions is much stronger than testing for equality of survival medians. Indeed, identical survival distributions imply equal median survival times, while the converse is not true.
Therefore, type I error is inflated when distributions differ between treatment groups for this method, which makes its application to real data limited. 

The BC test was modified by Tang and Jeong~\cite{tang2012median}, who employed contingency tables as an alternative to calculating the inverse of the BC test statistic in order to avoid estimating the density distribution of failure times. 
Rahbar et al.~\cite{rahbar2012nonparametric} proposed another nonparametric test, similar to the BC test, but where the difficulty of estimating the density is addressed by using the bootstrap approach to obtain the asymptotic variance of the test.
In general these nonparametric methods have inflated type I error rates, which make their use limited in practice, especially when the sample sizes are small~\cite{chen2016comparing}. 
Another extension of the BC test has been proposed by Chen and Zhang~\cite{chen2016comparing}.
However, like all aforementioned methods, this test can only be applied to compare one quantile at a time and authors do not propose an explicit calculation of the power of the test.
Moreover, these previous works do not address the formulation of clinical trials, which is a key aspect in the context of sample size estimation and in analyzing the impact of design parameters on the power of the test.

A versatile method for evaluating treatment effects by comparing pre-specified quantiles in each treatment group was proposed by Kosorok~\cite{kosorok1999two}. The author derived a nonparametric two-sample test for the comparison of quantiles that allows for testing the equality of multiple quantiles as well as testing the equality of one single quantile obtained at multiple analysis times. 
The test also applies to general censoring schemes, such as double censoring for instance, and several kinds of empirical distribution estimators. It can also be directly applied to group sequential clinical designs with staggered patient entry. 
\color{black}
Nevertheless, the practical application of this method is limited by the absence of explicit power formulas and minimal sample size calculations.
In this work, we fill this gap by deriving the necessary expressions, thereby extending Kosorok's method and enabling its use in clinical trial design.
\color{black}

\color{black}Moreover, \color{black} the estimation of the variance of the test statistic depends on the value of the density of survival at the quantile of interest. Kosorok proposed the use of kernel density estimators in order to complete this task. 
A drawback of this approach is that it requires the estimation of the density at all points and relies on an unknown bandwidth parameter, which has an impact on the estimator's performance~\cite{heidenreich2013bandwidth}.
To address this issue, we propose a resampling approach inspired by Lin et al.~\cite{lin2015conditional} to estimate the density directly at the point of interest without requiring a bandwidth parameter.
\color{black}
The full development of this procedure is presented in Farah et al.~\cite{farah2025note}, with additional details provided in the Supplemental Material. 
In that work, the performance of both density estimation methods was compared, showing that the resampling-based method outperformed the alternative in terms of mean squared error (MSE). 
\color{black}

Under Kosorok's framework, the key contribution of our work is the explicit derivation of closed-form formulas for the asymptotic power of the test for both univariate and multivariate quantile comparisons.
\color{black}
This advancement enables a novel approach to clinical trial planning and minimum sample size calculation.
Another highlight of our work is the refinement of the Kosorok test through a resampling-based method for density estimation.
\color{black}

This paper is organized as follows. First, we present the theoretical results for the univariate and multivariate tests of equality of quantiles. Next, we derive analytical power of the test in the context of planning clinical trials with known survival distributions. We then present applications of the method to a lung cancer dataset from the OAK randomized clinical trial~\cite{rittmeyer2017atezolizumab}, where we illustrate its use in comparing both single and multiple quantiles. 
We conclude with a discussion.

\section{Methods}
\label{sec:methods}
We consider a two-arms clinical trial, where $n_1$ patients are randomly allocated to treatment group 1 and $n_2$ to treatment group 2. 
We observe for each patient $i \in \{1,...,n_k\}$, $k=1,2$, the observed times $T_{ik} = \min(\tilde{T}_{ik}, C_{ik})$ and censoring status $\delta_{ik} = \mathbb{1}_{\tilde{T}_{ik} \leq C_{ik}}$, where $\tilde{T}_{ik}$ are the continuous times of interest and $\tilde{C}_{ik}$ the censoring times. 
Let $n=n_1+n_2$ and $\hat{\mu}_k=n_k/n$, we assume that $\hat{\mu}_k$ converges to an element $\mu_k$ in $(0,1)$. We also assume that the event times $\tilde{T}_{1k},\ldots,\tilde{T}_{n_kk}$ and the censoring times $C_{1k},\ldots,C_{n_kk}$ are independent, for $k=1,2$. Let $F_k$ be the cumulative distribution function \color{black} of failure times \color{black} in each treatment arm with $f_k$, $\Lambda_k$ and $S_k$ the density, cumulative hazard and survival functions, respectively. We denote the survival function of the observed times $T_{ik}$, as $H_k$. We define the usual inverse distributions as $F_{k}^{-1}(p) = \inf\{t: F_k(t) \geq p\}$, $k=1,2$, for a given probability $p \in (0,1)$. Following Kosorok~\cite{kosorok1999two}, we require the densities at the quantiles to be positive for both treatment groups. We also suppose that there exists $\epsilon > 0$ such that $H_k(F_{k}^{-1}(p) + \epsilon) > 0$, $k=1,2$. This condition is needed to ensure sufficient follow-up in order to be able to estimate each quantile of interest.

We denote as $\hat{F}_k$, $k=1,2$, the Kaplan-Meier estimator of $F_k$, from which we derive $\hat{F}_k^{-1}(p)$, the estimator of the inverse distributions at $p$. Our method is based on the asymptotic distribution of $\hat{F}_k^{-1}(p)$ as derived in Kosorok~\cite{kosorok1999two}. In this work, the asymptotic variance depends on the density at the quantile $F_{k}^{-1}(p)$, which needs to be estimated in order to construct a statistical test. This estimator is denoted by $\hat f_k$ which we propose to estimate by either kernel density estimation or a resampling procedure inspired by Lin et al.~\cite{lin2015conditional}
\color{black}
The resampling method, developed in Farah et al.~\cite{farah2025note}, consists of generating multiple realizations of the centered Gaussian variable with variance $\sigma_{\varepsilon}^2$, and then performing the least squares estimation which gives directly an estimation of the density at the quantile of interest $F_k^{-1}(p)$ for a given probability $p$ and group $k$. 
\color{black}
We propose an automatic grid-search algorithm in order to choose the variance of the generated Gaussian.
Our procedure has the advantage of providing an estimation of the density directly at the point of interest, contrary to kernel density estimation which requires an estimation at all data points.
Indeed, it is well known~\cite{silverman1986density} that kernel density estimators have a slow rate of convergence (in particular, slower than \color{black} the parametric rate \color{black} $n^{-1/2}$), that depends on the regularity of the density. For instance, if the density is assumed to be differentiable, then the rate of convergence is of order $n^{-1/3}$ for an optimal bandwidth of order $n^{-1/3}$.
\color{black}
On the other hand, estimating a parameter, such as the density at a single point, may allow for a faster rate of convergence.
\color{black}
Further details on the resampling and kernel density methods are provided in Supplemental Material \color{black} and in Farah et al.~\cite{farah2025note} \color{black}


\subsection{Univariate test}
We are interested in testing the null hypothesis $\mathcal{H}_0: F_1^{-1}(p) = F_2^{-1}(p)$ against the alternative hypothesis $\mathcal{H}_1: F_1^{-1}(p) - F_2^{-1}(p) = \Delta$, for a difference in quantiles $\Delta \in \mathbb{R}, \Delta \neq 0$. 
Such a null hypothesis allows to investigate the benefit, if any, of the experimental arm over the control arm at a given quantile.
Assuming the conditions outlined in Section \ref{sec:methods} hold for groups $k=1,2$, it follows from Lemma 1 from Kosorok~\cite{kosorok1999two}, with theoretical derivations detailed in Supplemental Material,
\begin{equation}
    \sqrt{n} (\hat{F}^{-1}_k(p) - F^{-1}_k(p) ) \xrightarrow{d} \mathcal{N}\left(0,(1-p)^2 \frac{\phi_k }{\mu_k f_k(F^{-1}_k(p))^2}\right) \text{ as } n\to \infty,
\end{equation}
where
\begin{equation}
    \phi_k = \int_0^{ F_k^{-1}(p)} \frac{d\Lambda_k(x)}{H_k(x)}\cdot
\end{equation}
We prove in Supplemental Material that under $\mathcal{H}_0$, 
\begin{equation}
    \sqrt{n}(\hat{F}_{1}^{-1}(p) - \hat{F}_{2}^{-1}(p) ) \xrightarrow{d} \mathcal{N}\left(0, \color{black}\sigma\color{black}^2 \right) \text{ as } n \to \infty,
\end{equation}
where
\begin{equation}
    \color{black}\sigma\color{black}^2 = (1-p)^2 \left( \dfrac{\phi_1}{\mu_1 f_1(F_{1}^{-1}(p))^2}+ \dfrac{\phi_2}{\mu_2 f_2(F_{2}^{-1}(p))^2} \right).
\end{equation}
We propose the following test statistic for the univariate test of equality of quantiles:
\begin{equation}
    \begin{aligned}
        \mathcal{T}_n &= \sqrt{n}\dfrac{\hat{F}_{1}^{-1}(p) - \hat{F}_{2}^{-1}(p) }{\color{black}\hat{\sigma}\color{black}}, \text{ with } \\
        \color{black}\hat{\sigma}\color{black}^2 &= (1-p)^2 \left( \dfrac{\hat{\phi_1}}{\hat{\mu}_1 \hat{f_1}(\hat{F}_{1}^{-1}(p))^2}+ \dfrac{\hat{\phi_2}}{\hat{\mu}_2\hat{f_2}(\hat{F}_{2}^{-1}(p))^2} \right),
\end{aligned}
\end{equation}
where $\hat{\phi_1}  $ and $\hat{\phi_2}$ are obtained by the usual Greenwood's estimate of the Kaplan-Meier variance, and $\hat{f}_1(\hat{F}_1^{-1}(p)), \hat{f}_2(\hat{F}_2^{-1}(p))$ are consistent estimators for the densities, such as kernel density estimators or the resampling procedure inspired by Lin et al.~\cite{lin2015conditional}

Then the following results hold:
\begin{enumerate}
    \item Under $\mathcal{H}_0 : F_1^{-1}(p) = F_2^{-1}(p)$, $\mathcal{T}_n \xrightarrow{d} \mathcal{N}(0,1)$ as $n \to \infty$.
    \item Under $\mathcal{H}_1 : F_1^{-1}(p) - F_2^{-1}(p) = \Delta$, $\mathcal{T}_n - \sqrt{n}\dfrac{\Delta}{\color{black}\hat{\sigma}\color{black}} \xrightarrow{d}\mathcal{N}(0,1) \text{ as } n \to \infty$.
\end{enumerate}
For a test with type I error $\alpha$ and power $1 - \beta$, as $n$ goes to infinity,
\begin{equation}
    P_{\mathcal{H}_0}(|\mathcal{T}_n| > q_{1 - \frac{\alpha}{2}}) \to \alpha,
\end{equation}
where $q_{1 - \frac{\alpha}{2}}$ is the quantile of order $1 - \alpha/2 $ of the standard normal distribution.

Let $\Phi$ be the cumulative distribution function of the standard normal distribution. From the asymptotic distribution of $\mathcal{T}_n$ under $\mathcal{H}_1$, we derive the following asymptotic formula for the power of the test:

\begin{equation}
    1 - \beta \approx 1 - \Phi \left( q_{1-\frac{\alpha}{2}} - \frac{\sqrt{n}}{\color{black}\hat{\sigma}\color{black}} \Delta \right) + \Phi\left(-q_{1-\frac{\alpha}{2}} - \frac{\sqrt{n}}{\color{black}\hat{\sigma}\color{black}} \Delta \right). 
    \label{power_univariate}
\end{equation}
Details and proofs can be found in Supplemental Material.

\subsection{Multivariate test}
We now present general results for multivariate two-sample tests of equality of quantiles.
The proofs are provided in Supplemental Material. 
We aim to test, for a given $J$, the null hypothesis $\mathcal{H}_0: F_1^{-1}(p_j) = F_2^{-1}(p_j)$, $j=1,...,J$ against the alternative hypothesis $\mathcal{H}_1: F_1^{-1}(p_j) - F_2^{-1}(p_j) = \Delta_j$,  $\exists j: \Delta_j \neq 0$.

Assuming the conditions outlined in section \ref{sec:methods} are satisfied, then for $k=1,2$:
\begin{equation}
    \color{black}
    \sqrt{n} \begin{bmatrix}
           \hat{F}_{k} ^{-1}(p_1) - F_k^{-1}(p_1)\\[1em]
           \vdots \\[1em]
           \hat{F}_{k} ^{-1}(p_J) - F_k^{-1}(p_J)
         \end{bmatrix} \xrightarrow{d} \mathcal{N}(0, \Upsilon_{F_k}) \text{ as } n \to \infty,
         \label{multi_F}
\end{equation}
where:
\begin{equation}
    (\Upsilon_{F_k})_{jl} = 
    \begin{cases}
        \dfrac{(1 - p_j)^2 \int_0^{F_k^{-1}(p_j)} \dfrac{d\Lambda_k(x)}{H_k(x)}}{\mu_k \left( f_k(F_k^{-1}(p_j)) \right)^2}, \text{if } j = l \\[2em]
        \dfrac{(1-p_j)(1-p_l) \int_0^{F_k^{-1}(p_j) \wedge F_k^{-1}(p_l)} \dfrac{d\Lambda_k(x)}{H_k(x)}}{\mu_k f_k(F_k^{-1}(p_j)) f_k(F_k^{-1}(p_l))}, \text{otherwise}.
    \end{cases}
\end{equation}
It follows that, under $\mathcal{H}_0$, 
    \begin{equation}
        \mathcal{Z}_n= \sqrt{n} \begin{bmatrix}
            \hat{F}_1^{-1}(p_1) - \hat{F}_2^{-1}(p_1)\\[1em]
            \vdots \\[1em]
            \hat{F}_1^{-1}(p_J) - \hat{F}_2^{-1}(p_J)
          \end{bmatrix} \xrightarrow{d} \mathcal{N}(0, \Psi) \text{ as } n \to \infty,
    \end{equation}
where $\Psi = \Upsilon_{F_1}  + \Upsilon_{F_2}$.
In order to implement the test, we assume $\Psi$ to be invertible.
For the multivariate test of equality of quantiles, our test statistic has the following form:
\begin{equation}
    \mathfrak{T}_n = \mathcal{Z}_n^T \hat{\Psi}^{-1} \mathcal{Z}_n,
\end{equation}
where $\hat{\Psi}$ is obtained by replacing each component by its estimate, as in the univariate case.

Then the following results hold:
\begin{enumerate}
    \item Under $\mathcal{H}_0: F_1^{-1}(p_j) = F_2^{-1}(p_j), j=1,...,J$, $\mathfrak{T}_n \xrightarrow{d} \chi_J^2$ as $n \to \infty$.
    \item Under $\mathcal{H}_1: F_1^{-1}(p_j) - F_2^{-1}(p_j) = \Delta_j, j = 1,..., J$, $\mathfrak{T}_n$ is asymptotically equivalent to $\chi_J^2(\Psi^{-1/2} \xi)$, an uncentered chi-squared distribution with $J$ degrees of freedom and mean $\Psi^{-1/2} \xi$, with 
    \begin{equation}
      \xi = \sqrt{n}\begin{bmatrix}
        \Delta_1\\[1em]
                \vdots \\[1em]
         \Delta_J      
     \end{bmatrix}\cdot
    \end{equation}
\end{enumerate}
The power of the multivariate test for equality of quantiles at level $\alpha$ as $n \to \infty$ expresses as:
\begin{equation}
    1 - \beta \approx 1- F_{\chi_J^2(\Psi^{-1/2} \xi)} (q_{J, 1-\alpha}),
    \label{stat_test_multi}
\end{equation}
where $F_{\chi_J^2(\Psi^{-1/2} \xi)}$ is the cumulative distribution function of the uncentered chi-squared distribution with $J$ degrees of freedom and mean $\Psi^{-1/2} \xi$, and $q_{J, 1-\alpha}$ denotes the quantile of order $1-\alpha$ of the chi-squared distribution with $J$ degrees of freedom.

\section{Illustration and results}
The proposed method for univariate and multivariate tests for quantile comparison is relevant in two main application frameworks.
First, it is useful in sample size planning scenarios where one is interested in designing a clinical trial to compare treatment effects in the quantile scale.
In this case, our method enables the calculation of statistical power for a given sample size at a fixed significance level. 
Similarly, it can be used to determine the minimum sample size required to achieve sufficient power for a test at a fixed level.
Second, the method can be applied when survival data from both control and experimental arms are available, and we are interested in testing the hypothesis of equality of quantiles between two survival distributions.
Both applications are presented in this section.
The code required to reproduce the results reported here was implemented in R and is publicly available at \url{https://github.com/beafarah/dens-estimation-at-quantile/}.

\subsection{Planning a clinical trial}
The results derived in the previous section allow sample size and power calculations for tests of equality of quantiles in the context of planning clinical trials in the presence of censoring. Assuming known distributions for survival times and censoring, one can compute the explicit power obtained by the test, or in an equivalent way, derive the minimum sample size required in order to achieve a fixed power.

We present simulations to illustrate the planning of a clinical trial using the proposed test of equality of quantiles. Two versatile scenarios were considered in the subsequent results, inspired by Eaton et al.~\cite{eaton2020designing}, which allow to illustrate multiple realistic frameworks including proportional and nonproportional survival (Figure \ref{fig:side_by_side_scenarios}).
Groups $k = 1,2$ correspond to the control and experimental arm, respectively.
In all scenarios, survival time in the control arm follows an exponential distribution with rate $\lambda_a$. 

The survival time distribution in the experimental arm is specified as follows, for each simulation scenario:
\begin{itemize}
    \item \underline{ Scenario 1 (proportional hazards):} Exponential with rate $\lambda_b$. 
    \item \underline{ Scenario 2 (late differences):} Piecewise exponential with rate $\lambda_a$ until time $t_{\text{cut}}$ and $\lambda_b$ onward.
\end{itemize}
In all scenarios, the distribution of censoring time is exponential with rate $\lambda_{\text{cens}}$. Using the expression derived in the previous section, it is possible to compute the analytical power as a function of the parameters for each scenario. 

Indeed, for a fixed difference in quantiles equal to $\Delta$, if both arms are exponential (scenario 1), one may deduce the expression for the rate of experimental arm as $\lambda_b = -\log(1-p)/(F_1^{-1}(p) - \Delta)$. Moreover, under independent censoring, it is possible to write the analytical expressions for the variance $\color{black}\sigma\color{black}^2$:
\begin{equation}
    \begin{aligned}
        \phi_1 &= \frac{\lambda_a}{\lambda_a + \lambda_{\text{cens}}} (e^{(\lambda_a+\lambda_{\text{cens}})F_1^{-1}(p)}-1) \\
        \phi_2 &= \frac{\lambda_b}{\lambda_b + \lambda_{\text{cens}}} (e^{(\lambda_b + \lambda_{\text{cens}})F_2^{-1}(p)}-1) \\
        \color{black}\sigma\color{black}^2 &= \dfrac{(1-p)^2 \frac{\lambda_a}{\lambda_a + \lambda_{\text{cens}}} (e^{(\lambda_a+\lambda_{\text{cens}})F_1^{-1}(p)}-1)}{\hat{\mu}_1 (\lambda_a e^{-\lambda_a F_1^{-1}(p)})^2}+ \dfrac{(1-p)^2 \frac{\lambda_b}{\lambda_b + \lambda_{\text{cens}}} (e^{(\lambda_b + \lambda_{\text{cens}})F_2^{-1}(p)}-1)}{\hat{\mu}_2(\lambda_b e^{-\lambda_b F_2^{-1}(p)})^2}\cdot 
    \end{aligned}
\end{equation}
These quantities allow us to have an explicit expression for the power of the test.

We derive similar results for the second scenario, where we observe nonproportional hazards and late treatment effects. 
We have the following expression for the quantile in the experimental arm:
\begin{equation}
    F_2^{-1}(p) = \begin{cases}
        -\frac{\log(1-p)}{\lambda_{a}}, 0 \leq p < 1-e^{-\lambda_{a} t_{\text{cut}}} \\
        t_{\text{cut}} - \left( \frac{\log(1-p) + \lambda_{a} t_{\text{cut}}}{\lambda_{b}}\right), p \geq 1-e^{-\lambda_{a} t_{\text{cut}}} 
    \end{cases}
    \label{eq:quantiles_piecewise}
\end{equation}
In order to ensure the realization of this scenario, we require that $F_1^{-1}(p) - t_{\text{cut}} > \Delta$, which is equivalent to $F_2^{-1}(p) > t_{\text{cut}}$ when specifying the parameters. 
For this scenario, we have the same $\phi_1$ as the one where both arms are exponential, while the expression for $\phi_2$ is now written as:
\begin{align}
    \phi_2 = &  \frac{\lambda_a}{\lambda_a + \lambda_{\text{cens}}} (e^{(\lambda_a + \lambda_{\text{cens}})  t_{\text{cut}}} - 1 ) + \left( \frac{\lambda_b}{\lambda_b + \lambda_{\text{cens}}} e^{(\lambda_a - \lambda_b)t_{\text{cut}}} \right)  
    \left( e^{(\lambda_b + \lambda_{\text{cens}}) F_2^{-1}(p)} - e^{(\lambda_b + \lambda_{\text{cens}}) t_{\text{cut}}} \right).
\end{align}
In all simulations we compare the medians in both groups and we assume that the number of patients in each arm is the same. 
We choose $\lambda_{\text{cens}}$ in order to have approximately 25\% of censoring in each group. We fix the rate of the control group in both scenarios as $\lambda_a = 1.5$ and $t_{\text{cut}} = 0.2$ in scenario 2.

\begin{figure}[ht]
    \centering
\begin{tikzpicture}
\begin{groupplot}[
    group style={group size=4 by 1, horizontal sep=1.1cm},
    width=4.2cm, height=4.2cm,
    xlabel={}, ylabel={},
    label style={font=\tiny},
    tick label style={font=\tiny},
    title style={font=\small},
    every axis/.append style={clip=false},
    every axis y label/.style={at={(axis description cs:-0.2,0.5)}, anchor=south, rotate=90, font=\scriptsize },
    every axis x label/.style={at={(axis description cs:0.5,-0.1)}, anchor=north, font=\scriptsize },
    every axis title/.style={at={(axis description cs:0.5,0.98)}, anchor=south, font=\small},
]

\nextgroupplot[xlabel={Time}, ylabel={Survival}, title={$\Delta = 0.1$},ytick={0,0.25,0.5,0.75,1.0} ]
\addplot[red, solid, thick] 
    table [x={"ts"}, y={"y1"}, col sep=space, header=true] {data_scenario1_delta01.txt};
\addplot[blue, dashed, thick] 
    table [x={"ts"}, y={"y2"}, col sep=space, header=true] {data_scenario1_delta01.txt};
\addplot[black, dashed, line width=0.5pt] coordinates {(0,0.5) (3,0.5)};
\node[below=7mm] at (current axis.south) {(a)};
\nextgroupplot[xlabel={Sample size}, ylabel={Power},title={$\Delta = 0.1$}, ytick={0.2, 0.4, 0.6, 0.8, 1}]
\addplot[orange, solid, line width=0.5pt] 
    table [x={"ns"}, y={"pwrs1"}, col sep=space, header=true] {power_scenario1_delta01.txt};
\addplot[black, dashed, line width=0.5pt] coordinates {(0,1) (1000,1)};
\node[below=7mm] at (current axis.south) {(b)};
\nextgroupplot[xlabel={Time}, ylabel={Survival},title={$\Delta = 0.2$}, ytick={0,0.25,0.5,0.75,1.0}]
\addplot[red, solid, thick] 
    table [x={"ts"}, y={"y1"}, col sep=space, header=true] {data_scenario1_delta02.txt};
\addplot[blue, dashed, thick] 
    table [x={"ts"}, y={"y2"}, col sep=space, header=true] {data_scenario1_delta02.txt};
\addplot[black, dashed, line width=0.5pt] coordinates {(0,0.5) (3,0.5)};
\node[below=7mm] at (current axis.south) {(c)};
\nextgroupplot[xlabel={Sample size},ylabel={Power}, title={$\Delta = 0.2$}, ytick={0.4, 0.6, 0.8, 1}]
\addplot[orange, solid, thick] 
    table [x={"ns"}, y={"pwrs1"}, col sep=space, header=true] {power_scenario1_delta02.txt};
\addplot[black, dashed, line width=0.5pt] coordinates {(0,1) (1000,1)};
\node[below=7mm] at (current axis.south) {(d)};
\end{groupplot}
\end{tikzpicture}
\raisebox{-2.0cm}{
\begin{tikzpicture}
\begin{groupplot}[
    group style={group size=4 by 1, horizontal sep=1.1cm},
    width=4.2cm, height=4.2cm,
    xlabel={}, ylabel={},
    label style={font=\small},
    tick label style={font=\tiny},
    title style={font=\small},
    every axis/.append style={clip=false},
    every axis y label/.style={at={(axis description cs:-0.2,0.5)}, anchor=south, rotate=90, font=\scriptsize },
    every axis x label/.style={at={(axis description cs:0.5,-0.1)}, anchor=north, font=\scriptsize },
    every axis title/.style={at={(axis description cs:0.5,0.98)}, anchor=south, font=\small},
]

\nextgroupplot[xlabel={Time}, ylabel={Survival}, title={$\Delta = 0.1$}, ytick={0,0.25,0.5,0.75,1.0}]
\addplot[red, solid, thick] 
    table [x={"ts"}, y={"y1"}, col sep=space, header=true] {data_scenario2_delta01.txt};
\addplot[blue, dashed, thick] 
    table [x={"ts"}, y={"y2"}, col sep=space, header=true] {data_scenario2_delta01.txt};
\addplot[black, dashed, line width=0.5pt] coordinates {(0,0.5) (3,0.5)};
\node[below=7mm] at (current axis.south) {(e)};
\nextgroupplot[xlabel={Sample size}, ylabel={Power}, title={$\Delta = 0.1$}]
\addplot[orange, solid, line width=0.5pt] 
    table [x={"ns"}, y={"pwrs1_piece"}, col sep=space, header=true] {power_scenario2_delta01.txt};
\addplot[black, dashed, line width=0.5pt] coordinates {(0,1) (1000,1)};
\node[below=7mm] at (current axis.south) {(f)};
\nextgroupplot[xlabel={Time},ylabel={Survival}, title={$\Delta = 0.2$}, ytick={0,0.25,0.5,0.75,1.0}]
\addplot[red, solid, thick] 
    table [x={"ts"}, y={"y1"}, col sep=space, header=true] {data_scenario2_delta02.txt};
\addplot[blue, dashed, thick] 
    table [x={"ts"}, y={"y2"}, col sep=space, header=true] {data_scenario2_delta02.txt};
\addplot[black, dashed, line width=0.5pt] coordinates {(0,0.5) (3,0.5)};
\node[below=7mm] at (current axis.south) {(g)};
\nextgroupplot[xlabel={Sample size},ylabel={Power}, title={$\Delta = 0.2$}] 
\addplot[orange, solid, thick] 
    table [x={"ns"}, y={"pwrs1_piece"}, col sep=space, header=true] {power_scenario2_delta02.txt};
\addplot[black, dashed, line width=0.5pt] coordinates {(0,1) (1000,1)};
\node[below=7mm] at (current axis.south) {(h)};
\end{groupplot}
\end{tikzpicture}
}

  \caption{Comparison of scenarios. Top row: Scenario 1 with true survival curves and power analyses. Bottom row: Scenario 2 with similar comparisons. The solid line represents the control arm and the dashed line represents the experimental arm.}
  \label{fig:side_by_side_scenarios}
\end{figure}

\newpage
For these two scenarios, we assess the performance of the power formula on finite samples. For this, ten thousand simulations were performed to compare the asymptotic power obtained by the explicit formula to the empirical power obtained from simulations. 
Results shown in Table \ref{T1} confirm that the analytical formula provides a good approximation even with modest sample sizes and that type I error is well controlled. 
\color{black}
As expected, increasing sample size yields empirical type I error and power values that are closer to the analytical ones.
\color{black}


        \begin{table}[h]
\small\sf\centering
{\color{black}
\begin{tabular}{lcc|cc}
\toprule
\multirow{2}{*}{\textbf{$\Delta$}} & \multicolumn{2}{c|}{\textbf{Scenario 1}} & \multicolumn{2}{c}{\textbf{Scenario 2}} \\
\cmidrule{2-5}
& \textbf{Empirical} & \textbf{Formula}  & \textbf{Empirical} & \textbf{Formula}  \\
\midrule
\multicolumn{5}{c}{\textbf{Sample size $n_i = 50$}} \\
\midrule
0   & 0.0103 & 0.05    & 0.0015  & 0.05   \\
0.1 & 0.0504 & 0.1236 & 0.0174  & 0.1153 \\  
0.2 & 0.3507 & 0.4147  & 0.3908  & 0.4682 \\
\midrule
\multicolumn{5}{c}{\textbf{Sample size $n_i = 100$}} \\
\midrule
0   & 0.0257 & 0.05   & 0.0060  & 0.05    \\
0.1 & 0.1583 & 0.2000  & 0.0991  & 0.1831  \\
0.2 & 0.7280 & 0.6939  & 0.8645  & 0.7577 \\
\midrule
\multicolumn{5}{c}{\textbf{Sample size $n_i = 500$}} \\
\midrule
0   & 0.047  & 0.05   & 0.047 & 0.05    \\
0.1 & 0.714  & 0.703   & 0.782  & 0.766   \\
0.2 & 1.000  & 1.000   & 1.000  & 1.000  \\
\bottomrule
\end{tabular}
}
\caption{Type I error and power of the test of equality of quantiles for sample sizes $n_i=50,100,500$, at $p = 0.5$.}
\label{T1}
\end{table}

          One interest of the explicit power formula is its ability to compute the minimum sample size required in order to detect a fixed difference at a desired power. We illustrate this application by fixing a desired power and treatment effect and comparing the minimum sample size required for each scenario when testing for the equality of medians. The results are provided in Table \ref{tab:power_delta_patients}.
          \begin{table}[h]
            \small\sf\centering
            \begin{tabular}{cccc}
            \toprule
            \multirow{2}{*}{\textbf{Power}} & \multirow{2}{*}{\textbf{$\Delta$}} & \textbf{Scenario 1} & \textbf{Scenario 2} \\
            \cmidrule{3-4}
             & & \textbf{Sample Size} & \textbf{Sample Size} \\
            \midrule
            \multirow{2}{*}{0.95}           & 0.1 & 1047 & 901 \\
                                            & 0.2 & 214  & 173 \\
            \midrule
            \multirow{2}{*}{0.90}           & 0.1 & 846  & 729 \\
                                            & 0.2 & 173  & 140 \\
            \midrule
            \multirow{2}{*}{0.80}           & 0.1 & 632  & 545 \\
                                            & 0.2 & 129  & 105 \\
            \bottomrule
            \end{tabular}
            \caption{Minimal sample size per group.\label{tab:power_delta_patients}}
            \end{table}

            Analytical power can be plotted at increasing quantile difference for various sample sizes, which is presented in Figure \ref{fig:power_deltas}. 
            Across both scenarios, power increases with sample size, as expected. 
            Differences between treatment arms at the median are illustrated in scenario 2 for $\Delta \in [0,0.25]$ in order to satisfy the condition $F_1^{-1}(p) - t_{\text{cut}} > \Delta$.

\begin{figure}
\centering

\begin{minipage}{0.48\textwidth}
\centering
\begin{tikzpicture}
\begin{axis}[
    width=1.1\textwidth, height=5cm,
    xlabel={$\Delta$},
    ylabel={Power},
    tick label style={font=\scriptsize},
    label style={font=\small},
    legend style={font=\tiny, at={(1,0)}, anchor=south east,
    nodes={inner sep=2pt},       
    row sep=0pt,                 
    column sep=2pt,              
    /tikz/inner xsep=1pt,        
    /tikz/inner ysep=1pt,},
    grid=none
]

\addlegendimage{empty legend}
\addlegendentry{\textbf{$n_i$}} 

\addplot[vir1, dashed, thick] table [x={"deltas"}, y={"power_deltas1"}, col sep=space, header=true] {power_deltas_scenario1.txt};
\addlegendentry{$50$}
\addplot[vir2, dashed, thick] table [x={"deltas"}, y={"power_deltas2"}, col sep=space, header=true] {power_deltas_scenario1.txt};
\addlegendentry{$100$}
\addplot[vir3, dashed, thick] table [x={"deltas"}, y={"power_deltas3"}, col sep=space, header=true] {power_deltas_scenario1.txt};
\addlegendentry{$300$}
\addplot[vir4, dashed, thick] table [x={"deltas"}, y={"power_deltas4"}, col sep=space, header=true] {power_deltas_scenario1.txt};
\addlegendentry{$500$}
\addplot[vir5, dashed, thick] table [x={"deltas"}, y={"power_deltas5"}, col sep=space, header=true] {power_deltas_scenario1.txt};
\addlegendentry{$1000$}
\end{axis}
\end{tikzpicture}
\end{minipage}
\hspace{0.1cm} 
\begin{minipage}{0.48\textwidth}
\centering
\begin{tikzpicture}
\begin{axis}[
    width=1.1\textwidth, height=5cm,
    xlabel={$\Delta$},
    xtick = {0, 0.05, 0.1, 0.15, 0.2, 0.25},
    scaled ticks=false, 
        tick label style={
        /pgf/number format/fixed,           
        /pgf/number format/precision=2      
    },
    tick label style={font=\scriptsize},
    label style={font=\small},
    legend style={font=\scriptsize, at={(1,1)}, anchor=north east},
    grid=none
]
\addplot[vir1, dashed, thick] table [x={"deltas.deltas...last_delta."}, y={"power_deltas1"}, col sep=space, header=true] {power_deltas_scenario2.txt};
\addplot[vir2, dashed, thick] table [x={"deltas.deltas...last_delta."}, y={"power_deltas2"}, col sep=space, header=true] {power_deltas_scenario2.txt};
\addplot[vir3, dashed, thick] table [x={"deltas.deltas...last_delta."}, y={"power_deltas3"}, col sep=space, header=true] {power_deltas_scenario2.txt};
\addplot[vir4, dashed, thick] table [x={"deltas.deltas...last_delta."}, y={"power_deltas4"}, col sep=space, header=true] {power_deltas_scenario2.txt};
\addplot[vir5, dashed, thick] table [x={"deltas.deltas...last_delta."}, y={"power_deltas5"}, col sep=space, header=true] {power_deltas_scenario2.txt};
\end{axis}
\end{tikzpicture}
\end{minipage}

                \caption{Power for various differences in quantiles in scenarios 1 (on the left) and 2 (on the right).}
                \label{fig:power_deltas}
\end{figure}

            \color{black}
            The Supplemental Material provides additional simulation results, in particular for more extreme quantiles and higher censoring rates, along with the corresponding computational times for all scenarios.
            \color{black}
\FloatBarrier
\subsection{Application of the test on data from the OAK study}
\begin{figure}[h]
  \centering
  \includegraphics[scale = .45]{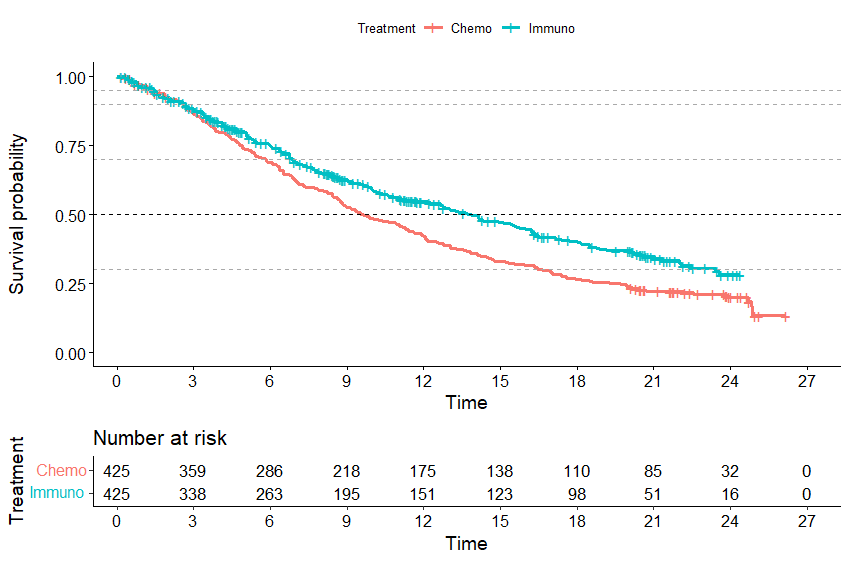}
  \caption{Reconstructed Kaplan-Meier curves. The dashed lines represent the survival quantiles at probabilities 0.05, 0.1, 0.3, 0.5 and 0.7.}
  \label{fig:combined_km}
\end{figure}

\color{black}
In this section, we apply the test of equality of quantiles to the OAK randomized phase III clinical trial, registered with ClinicalTrials.gov, number NCT02008227. 
This study was conducted to evaluate the efficacy and safety of the immunotherapy treatment atezolizumab compared with the standard of care chemotherapy docetaxel in patients with previously treated metastatic non-small-cell lung cancer.
The primary efficacy analysis was performed in the first 850 of 1225 enrolled patients, recruited between March 11, 2014 and April 29, 2015.
The primary endpoint of this multicentre randomized controlled trial was the overall survival, with approximately 30\% of censored of observations at the time of analysis.
\color{black}
Following Mboup et al.~\cite{mboup2021insights}, we use reconstructed survival data generated by the algorithm developed in Rittmeyer et al.~\cite{rittmeyer2017atezolizumab} to emulate survival times for both treatment arms, enabling our analysis.
The reconstructed Kaplan-Meier curves are seen in Figure \ref{fig:combined_km}.

The implementation of the test statistic requires the value of the density at the quantiles for both treatment arms.
\color{black}
To this end, two approaches for density estimation are compared: kernel density estimation and our resampling procedure inspired by Lin et al.~\cite{lin2015conditional}, here referred to as KDE and LS respectively.
The KDE method is described in  F{\"o}ldes et al.~\cite{foldes1981strong} and Diehl and Stute~\cite{diehl1988kernel}, with additional details provided in the Supplemental Material.
The LS method, presented in Farah et al.~\cite{farah2025note}, consists of generating multiple samples of the zero-mean Gaussian variable with variance $\sigma^2_{\varepsilon}$, from which the density at the quantile is directly estimated using least squares estimation.
In practical applications, $\sigma^2_{\varepsilon}$ is selected through a grid-search algorithm detailed in Farah et al.~\cite{farah2025note}.
In our implementation with the OAK clinical trial data, we draw $10^3$ zero-mean Gaussians, considering $\sigma_{\varepsilon}$ values ranging from $0.1$ to $10$ in increments of $0.05$.
For the KDE method, the bandwidth parameter for the KDE is selected via leave-one-out cross-validation over candidate values ranging from $0.1$ to $1$ in increments of $0.02$, with a Gaussian kernel.
\color{black}
\subsubsection{Application of the univariate test}
\vspace{\baselineskip}
We compute the power of the univariate test of equality of quantiles for OAK data. 
The test was applied at quantiles of order 0.3, 0.5 and 0.7, and differences between treatment arms are estimated for each quantile. 
The results are presented in Table \ref{results_univariate}.
\color{black} The values for the standard error of the resampled Gaussians selected by grid-search for the LS method are equal to $ 2.5, 1.9$ and $2.3$, respectively. \color{black}

For all quantiles considered, patients treated with immunotherapy have a positive survival average benefit compared with patients treated with immunotherapy, which can be directly interpreted in terms of survival time. 
Taking the median, for instance, we may say that, on average, patients from the immunotherapy group have 50\% of chance to survive 4.04 months more than patients in the chemotherapy group.

\color{black}
At the median, the test of equality of medians proposed in Tang and Jeong~\cite{tang2012median} yields a $p$-value of $2.22 \times 10^{-3}$ for this setting.
By comparison, the LS method provides stronger statistical evidence at the median, while also being directly applicable to arbitrary quantiles and accommodating multivariate quantile inference, in contrast to the approach of Tang and Jeong.
\color{black}

\begin{table}[h]
  \small\sf\centering
  \begin{tabular}{cc|cc|cc}
  \toprule
  \multirow{2}{*}{\textbf{$p$}} & \multirow{2}{*}{\textbf{$\hat{\Delta}$}} & \multicolumn{2}{c|}{\textbf{$p$-value}} & \multicolumn{2}{c}{\textbf{Test statistic}} \\
  \cmidrule{3-6}
   & & \textbf{$\textsc{LS}$} & \textbf{KDE} & \textbf{$\textsc{LS}$} & \textbf{KDE} \\
  \midrule
  0.3 & -1.01 & $1.13 \times 10^{-2}$ & $8.49\times 10^{-2}$ & -2.53 & -1.72 \\
  0.5 & -4.04 & $5.03\times 10^{-4}$ & $5.35\times 10^{-3}$ & -3.48 & -2.79 \\
  0.7 & -6.76 & $4.39 \times 10^{-9}$ & $5.05\times 10^{-7}$ & -5.87 & -5.02 \\
  \bottomrule
  \end{tabular}
  \caption{$p$-values and test statistics for $\textsc{LS}$ and KDE methods for the univariate test. \label{tab:pvalues_teststats}}
  \label{results_univariate}
  \end{table}

All methods lead to the same statistical conclusion of rejection of the null hypothesis at each of the tested quantiles.
In all cases, the test with the LS method for density estimation yields the most significant $p$-values.

\subsubsection{Application of the multivariate test}

One advantage of the proposed test of equality of quantiles is its direct application for multivariate two-sample tests of equality of quantiles. 
In this section we present the results obtained for the comparison of pairs of quantiles, which can be generalized in order to compare any set of $J$ quantiles.
We apply the multivariate test to three pairs of quantiles: \{(0.05, 0.1), (0.1, 0.5), (0.5,0.7)\}. 
The results are provided in Table \ref{tab:multi_test}, where either KDE or LS were used for computing the density at the quantiles which are needed for the estimation of the matrix of variance-covariance of the test statistic. 
\color{black}The standard error values selected in the LS method are $2.55, 2.3,$ and $1.5$, corresponding respectively to the three considered pairs of quantiles.\color{black}

\begin{table}[h]
  \small\sf\centering
  \begin{tabular}{cc|cc|cc}
  \toprule
  \multirow{2}{*}{\textbf{$p$}} & \multirow{2}{*}{\textbf{$\hat{\Delta}$}} & \multicolumn{2}{c}{\textbf{$p$-value}} & \multicolumn{2}{c}{\textbf{Test statistic}} \\
  \cmidrule{3-6}
   & & \textbf{$\textsc{LS}$} & \textbf{KDE} & \textbf{$\textsc{LS}$} & \textbf{KDE} \\
  \midrule
  0.05, 0.1 & -0.03, -0.06 & $9.83 \times 10^{-1}$ & $9.91\times 10^{-1}$ & 0.03 & 0.02 \\
  0.1, 0.5 & -0.06, -4.04 & $3.39\times 10^{-2}$ & $1.18\times 10^{-1}$ & 6.77 & 4.28 \\
  0.5, 0.7 & -4.04, -6.76 & $4.61\times 10^{-6}$ & $1.80\times 10^{-3}$ & 24.58 & 12.65 \\
  \bottomrule
  \end{tabular}
  \caption{$p$-values and test statistics for $\textsc{LS}$ and KDE methods for the multivariate test. \label{tab:multi_test}}
  
  \end{table}

The first row of Table \ref{tab:multi_test} corresponds to the scenario where the null hypothesis is satisfied. 
In this case, the null hypothesis is not rejected by both methods with $p$-values close to 1.
In the second row, we have a case where the difference of quantiles is close to zero (equal to 0.06 in absolute value) for one of the probabilities and close to 4 in absolute value for the other.
For a test of level 0.05, one rejects the null hypothesis when using the LS method for density estimation and does not reject it when the density is estimated by the KDE procedure.
This is the only case among the three scenarios where, depending on the density estimation procedure, the conclusion of the test changes.
This suggests a loss of power for the KDE procedure, which might be explained by the increased variability in this method due to its need to estimate the density at all points.

Lastly, we explore a scenario where there is a marked difference in quantiles for both probabilities, equal in absolute value to 4.04 and 6.76 for the survival quantiles 0.5 and 0.7 respectively.
In this case, both methods reject the null hypothesis, and \color{black} the test using the LS density estimation yields the most significant $p-$value.\color{black}

\color{black}
          The multivariate test of equality of quantiles assesses whether multiple quantiles are simultaneously equal across treatment groups at a given time.
          When the null hypothesis of equality across all quantiles is rejected, it is often of interest to determine which specific quantiles account for the differences.
          In this case, following a reviewer’s suggestion, we complement the global test with separate univariate tests for each quantile, adjusted using the Bonferroni correction.
          We illustrate this approach with the multivariate test applied to the quantiles $\{0.05, 0.1, 0.5, 0.7\}$. 
          The grid-search algorithm, using the same grid as previously described, selects $\sigma^2_{\varepsilon} = 2.5$.
          The resulting test statistic is equal to $24.66$, which yields a $p$-value of $5.87 \times 10^{-5}$, providing strong evidence against the null hypothesis of equal quantiles.
         To identify which quantiles differ between groups, we conduct univariate tests with Bonferroni correction.
         The standard deviations obtained from grid search for quantiles $0.05$, $0.1$, $0.5$, and $0.7$ are respectively equal to $2.3, 1.65, 1.9,$ and $2.3$.
         The corresponding adjusted $p-$values are $8.77 \times 10^{-1}$, $7.94 \times 10^{-1}$, $4.44\times 10^{-4}$, and $5.80 \times 10^{-9}$.
         These results indicate that the null hypothesis is not rejected for the lower quantiles ($0.05$ and $0.1$), but it is strongly rejected for the median ($0.5$) and upper quantile ($0.7$).
        The computational runtime for the multivariate test is approximately $24.21$ seconds, while the total runtime for all four univariate tests combined is about $27.43$ seconds.
          The proposed strategy offers a practical guideline, whereby the multivariate test evaluates whether global differences exist across quantiles, and the adjusted univariate tests localize those differences.
\color{black}

\section{Discussion}
In this paper we derived the analytical expressions for the power calculation for the univariate and multivariate tests of equality of quantiles proposed by Kosorok~\cite{kosorok1999two}.
The explicit formulas can be used in different situations when evaluating the effects of a new treatment against standard of care, such as computing minimum sample size when planning clinical trials or comparing quantiles in the presence of clinical data.
The power formula was analytically derived and its asymptotic behavior was studied in simulations for the scenarios where there are proportional and nonproportional hazards. 
The power formula can be derived for general situations, by assuming different distributions for survival and censoring.
The proposed test was illustrated in the context of immuno-oncology trials with late treatment effects and nonproportional hazards. 
Estimation techniques are needed in order to perform the necessary power calculations for the test in presence of data. We showed the results obtained by using kernel density estimation as well as a resampling technique proposed \color{black}in Farah et al.\cite{farah2025note} \color{black}, with an improved power for the resampling approach.
\color{black}
Of note, the choice of quantiles for the test should be guided by clinical relevance and data availability, in order to ensure reliable estimation. In particular, high quantiles might not be reached or might be poorly estimated due to tail estimation problems if only few events are observed at late times.
\color{black}

One major advantage of our proposed method is that it allows for multivariate tests, which can be further extended to group sequential clinical trials with staggered entry of patients and several interim analyses. 
Furthermore, although the proposed test is designed for detecting the differences in quantiles of survival times, one could extend \color{black} the power formulas in a co-primary endpoints context, where the goal is to \color{black} assess the effect of a randomized treatment jointly on the hazard and a given quantile or on the restricted mean survival time (RMST)~\cite{eaton2020designing} and a given quantile. 
Indeed, there is a subtle connection between RMST and percentile of survival because in practice, the choice of a clinically relevant restriction time imposes a percentile of survival. This is left to future research work. 



\section*{\normalsize Funding}
The author(s) disclosed receipt of the following financial support for the research, authorship, and/or publication of this article: Research in this article was supported by Ecole doctorale de Santé Publique, Université Paris-Saclay and Université de Versailles-Saint-Quentin-en-Yvelines.

\section*{\normalsize Declaration of conflicting interests}
The author(s) declared no potential conflicts of interest with respect to the research, authorship, and/or publication of this article.

\section*{\normalsize Code availability}
All data processing and statistical analysis were performed with the R statistical computing software version 4.3.2. 
Source code is available at \url{https://github.com/beafarah/dens-estimation-at-quantile/}.

\section*{\normalsize Supplemental Material}
Supplemental Material for this article is available online.

\bibliographystyle{vancouver}
\bibliography{biblio}

\end{document}


\runninghead{Anonymous}

\title{Supplemental material: Designing clinical trials for the comparison of single and multiple quantiles with right-censored data}

\author{Beatriz Farah\affilnum{1,2,3},Olivier Bouaziz\affilnum{2,4}\textsuperscript{*} and Aurélien Latouche\affilnum{1,5}\textsuperscript{*}}

\affiliation{%
\affilnum{1}INSERM U1331, Institut Curie, Saint-Cloud, France\\
\affilnum{2}CNRS, MAP5, Université Paris Cité, Paris, France\\
\affilnum{3}Université Paris-Saclay, UVSQ, Versailles, France\\
\affilnum{4}CNRS, UMR 8524 - Laboratoire Paul Painlevé, Univ. Lille, Lille, France\\
\affilnum{5}Conservatoire National des Arts et Métiers, Paris, France%
}


\corrauth{Beatriz Farah, INSERM U1331, Institut Curie, Mines Paris Tech, Paris/Saint-Cloud, France.\\ *These authors contributed equally to this work.}

\email{beatriz.farah@math.cnrs.fr}

\maketitle

\tableofcontents

\newpage
\section{Proofs and Technical details}
\subsection{Univariate test}

In this section, we present the proofs for the results in Section~2.1 of the main document, for the univariate test. We first recall the expression of the statistical test in the univariate case.

\begin{equation}
    \begin{aligned}
        \mathcal{T}_n &= \sqrt{n}\dfrac{\hat{F}_{1}^{-1}(p) - \hat{F}_{2}^{-1}(p) }{\hat{\sigma}}, \text{ with } \\
        \hat{\sigma}^2 &= (1-p)^2 \left( \dfrac{\hat{\phi_1}}{\hat{\mu}_1 \hat{f_1}(\hat{F}_{1}^{-1}(p))^2}+ \dfrac{\hat{\phi_2}}{\hat{\mu}_2\hat{f_2}(\hat{F}_{2}^{-1}(p))^2} \right).
\end{aligned}
\end{equation}

\begin{result}
    Under $\mathcal{H}_0 : F_1^{-1}(p) = F_2^{-1}(p)$, $\mathcal{T}_n \xrightarrow{d} \mathcal{N}(0,1)$ as $n \to \infty$
\end{result}
\begin{proof}
    From Lemma 1 from Kosorok~\cite{kosorok1999two}, for $k=1,2$,
    \begin{equation}
        \hat{F}^{-1}_k(p) - F^{-1}_k(p) = \frac{\hat{F}_k(F^{-1}_k(p)) - p}{f_k(F^{-1}_k(p))} + o_p\left(\frac{1}{\sqrt{n_k}} \right)
    \end{equation}
    It follows from Theorem 6.3.1 from Fleming and Harrington~\cite{1} that:
    \begin{equation}
        \sqrt{n_k} \left(\frac{\hat{F}_k(F^{-1}_k(p)) - p }{f_k(F^{-1}_k(p))}\right) 
        \xrightarrow{d}  \mathcal{N}\left(0, (1-p)^2\frac{\phi_k}{f_k(F_k^{-1}(p))} \right), \text{ with}
    \end{equation}
    \begin{equation}
        \phi_k = \int_0^{ F_k^{-1}(p)} \frac{d\Lambda_k(x)}{H_k(x)}\cdot
    \end{equation}
    From the independence between treatment groups, as $n \to \infty$, we have that
    \begin{equation}
        \begin{aligned}
            \sqrt{n} \bigg[ 
                \frac{\hat{F}_{1}(F_1^{-1}(p)) - p }{f_1(F_1^{-1}(p))} 
                - \frac{\hat{F}_{2}(F_2^{-1}(p)) - p }{f_2(F_2^{-1}(p))} 
            \bigg] 
            &\xrightarrow{d} 
            \mathcal{N} \bigg( 
                0, \, (1-p)^2 \bigg[ 
                    \frac{\phi_1}{\mu_1 f_1(F_1^{-1}(p))^2} 
                    + \frac{\phi_2}{\mu_2 f_2(F_2^{-1}(p))^2} 
                \bigg] 
            \bigg).
        \end{aligned}
        \label{eq:sigma_H0}
    \end{equation}
    Then, $\hat{\sigma} \to \sigma$ as $n \to \infty$ from the consistency of Greenwood's estimator of the Kaplan-Meier variance, the consistency of the density estimators and the consistency of $\hat\mu_k$, $k=1,2$. 
    Finally under $\mathcal{H}_0$, we have the decomposition 
    \begin{equation}
        \hat{F}^{-1}_1(p) -  \hat{F}^{-1}_2(p) = \hat{F}^{-1}_1(p) - F^{-1}_1(p) -  (\hat{F}^{-1}_2(p) - F^{-1}_2(p)),
    \end{equation}
    which concludes the proof.
\end{proof}
\begin{result}
    Under $\mathcal{H}_1 : F_1^{-1}(p) - F_2^{-1}(p) = \Delta$, $\mathcal{T}_n - \sqrt{n}\dfrac{\Delta}{\hat{\sigma}} \xrightarrow{d}\mathcal{N}(0,1) \text{ as } n \to \infty$
\end{result}
\begin{proof}
    Under $\mathcal{H}_1$, 
\begin{equation}
    \begin{aligned}
        \sqrt{n}(\hat{F}_{1}^{-1}(p) - \hat{F}_{2}^{-1}(p)) &= \sqrt{n}(\hat{F}_{1}^{-1}(p) - F_1^{-1}(p)) - \sqrt{n}( \hat{F}_{2}^{-1}(p) - F_2^{-1}(p)) + \sqrt{n} \Delta. 
    \end{aligned}
\end{equation}
We denote 
\begin{equation}
    Z_n = \sqrt{n}(\hat{F}_{1}^{-1}(p) - F_1^{-1}(p)) - \sqrt{n}( \hat{F}_{2}^{-1}(p) - F_2^{-1}(p)).
\end{equation}
Then, from Lemma 1 from Kosorok~\cite{kosorok1999two}, 
\begin{equation}
    Z_n = \sqrt{n} \left(\frac{\hat{F}_1(F_1^{-1}(p)) - p}{f_1(F_1^{-1}(p))} \right) - \sqrt{n} \left(\frac{\hat{F}_2(F_2^{-1}(p)) - p}{f_2(F_2^{-1}(p))} \right) + o_p\left(\frac{1}{\sqrt{n}}\right). 
\end{equation}
It follows, for $n \to \infty$:
\begin{equation}
    Z_n \xrightarrow{d} \mathcal{N}\left(0, \frac{(1-p)^2 \phi_1}{\mu_1 f_1(F_1^{-1}(p))^2} + \frac{(1-p)^2 \phi_2}{\mu_2 f_2(F_2^{-1}(p))^2} \right)
\end{equation}
From the consistency of $\hat\sigma$, we therefore have $Z_n/\hat\sigma \xrightarrow{d} \mathcal{N}(0,1)$, which gives the desired result using the relation 
\begin{equation}
    \mathcal{T}_n  = Z_n/\hat \sigma + \sqrt{n} \Delta/\hat \sigma.
\end{equation}
\end{proof}

\begin{result}
    We have the following asymptotic formula for the power of the test of level $\alpha$:
\begin{equation}
    1 - \beta \approx 1 - \Phi \left( q_{1-\alpha/2} - \frac{\sqrt{n}}{\hat{\sigma}} \Delta \right) + \Phi\left(-q_{1-\alpha/2} - \frac{\sqrt{n}}{\hat{\sigma}} \Delta \right)
    \label{power_univariate}
\end{equation}
where $\Phi$ is the cumulative distribution function of the standard normal variable and $q_{1-\alpha/2}$ is the quantile of level $1-\alpha/2$ of the standard normal distribution.
\end{result}

\begin{proof}
    Using the derived expression for the test statistic and denoting as $q_{1-\frac{\alpha}{2}}$ the quantile of order $1- \frac{\alpha}{2}$ of the standard normal distribution,
\begin{equation}
    \begin{aligned}
        1 - \beta &= P_{\mathcal{H}_1}(|\mathcal{T}_n| > q_{1-\frac{\alpha}{2}}) \\
         &= P_{\mathcal{H}_1}\left(\sqrt{n} \left|\frac{\hat{F}_{1}^{-1}(p) - \hat{F}_{2}^{-1}(p) }{\hat{\sigma}} \right| > q_{1-\frac{\alpha}{2}}\right) \\
         &= P_{\mathcal{H}_1}\left( \sqrt{n} \left|\left(\frac{\hat{F}_{1}^{-1}(p) - F_1^{-1}(p)}{\hat{\sigma}}\right) - \left( \frac{\hat{F}_{2}^{-1}(p) - F_2^{-1}(p)}{\hat{\sigma}}\right) + \frac{\Delta}{\hat{\sigma}}\right| > q_{1-\alpha/2}\right) \\
         &\approx P_{\mathcal{H}_1}\left( \frac{\sqrt{n}}{\hat\sigma} \left(\frac{\hat{F}_{1}(F_1^{-1}(p)) -p}{f_1(F_1^{-1}(p))} -  \frac{\hat{F}_{2}(F_2^{-1}(p)) - p}{f_2(F_2^{-1}(p))}\right)  > q_{1-\alpha/2} - \frac{\sqrt{n} \Delta}{\hat{\sigma}} \right) \\
         &+ P_{\mathcal{H}_1}\left( \frac{\sqrt{n}}{\hat\sigma} \left(\frac{\hat{F}_{1}(F_1^{-1}(p)) -p}{f_1(F_1^{-1}(p))} -  \frac{\hat{F}_{2}(F_2^{-1}(p)) - p}{f_2(F_2^{-1}(p))}\right)  < -q_{1-\alpha/2} - \frac{\sqrt{n} \Delta}{\hat{\sigma}} \right) \\
         &\approx P_{\mathcal{H}_1}\left(\frac{Z_n}{\hat\sigma}   > q_{1-\alpha/2} - \frac{\sqrt{n} \Delta}{\hat{\sigma}} \right) + P_{\mathcal H_1}\left(\frac{Z_n}{\hat\sigma}  < -q_{1-\alpha/2} - \frac{\sqrt{n} \Delta}{\hat{\sigma}}\right),
    \end{aligned}
\end{equation}
where the expression in the fourth equality comes from Lemma 1 from Kosorok~\cite{kosorok1999two}. We conclude from the convergence of $Z_n/\hat\sigma$ towards a centered Gaussian random variable.
\end{proof}

\subsection{Multivariate test}
In this section, we present the proofs for the results in Section~2.2 of the main document, for the multivariate test. 

\begin{result} For $k=1,2$,
\begin{equation*}
        \sqrt{n} \begin{bmatrix}
               \hat{F}_{k} ^{-1}(p_1) - F_k^{-1}(p_1)\\[1em]
               \vdots \\[1em]
               \hat{F}_{k} ^{-1}(p_J) - F_k^{-1}(p_J)
             \end{bmatrix} \xrightarrow{d} \mathcal{N}(0, \Upsilon_{F_k}) \text{ as } n \to \infty,
             \label{multi_F}
    \end{equation*}

    where:
    \begin{equation*}
        (\Upsilon_{F_k})_{jl} = \begin{cases}
            \dfrac{(1 - p_j)^2 \int_0^{F_k^{-1}(p_j)} \dfrac{d\Lambda_k(x)}{H_k(x)}}{\mu_k (f_k(F_k^{-1}(p_j)))^2}, \text{ if } j = l \\[2em]
            \dfrac{(1-p_j)(1-p_k)  \int_0^{F_k^{-1}(p_j) \wedge F_k^{-1}(p_l)} \dfrac{d\Lambda_k(x)}{H_k(x)} }{ \mu_k f_k(F_k^{-1}(p_j)) f_k(F_k^{-1}(p_l))} \text{, otherwise. }
        \end{cases}
    \end{equation*}
\end{result}
\begin{proof}
    From Lemma 1 in Kosorok~\cite{kosorok1999two}, we have:
\begin{align}
    \begin{bmatrix}
           \hat{F}_k^{-1}(p_1) - F_k^{-1}(p_1)\\[1em]
           \vdots \\[1em]
           \hat{F}_k^{-1}(p_J) - F_k^{-1}(p_J)
         \end{bmatrix} &= 
       U_k + o_p(1/\sqrt{n_k}),
\end{align}
where
\begin{equation}
    U_{k} =          \begin{bmatrix}
        \dfrac{\hat{F}_k(F_k^{-1}(p_1)) - p_1}{f_k(F_k^{-1}(p_1))}\\[2em]
        \vdots \\[1em]
        \dfrac{\hat{F}_k(F_k^{-1}(p_J)) - p_J}{f_k(F_k^{-1}(p_J))}
      \end{bmatrix}\cdot
\end{equation}
From Theorem 6.3.1 from Fleming and Harrington~\cite{1}, $\sqrt{n_k}U_{k}$ converges to a centered multivariate Gaussian random variable where its variance matrix has entries $(j,l)$ equal to:
\begin{equation}
    \dfrac{(1-p_j)(1-p_l)  \int_0^{F_k^{-1}(p_j) \wedge F_k^{-1}(p_l)} \dfrac{d\Lambda_k(x)}{H_k(x)} }{  f_k(F_k^{-1}(p_j)) f_k(F_k^{-1}(p_l))}\cdot
\end{equation}
The result follows since $n_k/n \to \mu_k$ as $n \to \infty$.

\end{proof}
\begin{result} Under $\mathcal{H}_0: F_1^{-1}(p_j) = F_2^{-1}(p_j), j=1,...,J$, 
    \begin{equation*}
        \mathcal{Z}_n= \sqrt{n} \begin{bmatrix}
            \hat{F}_1^{-1}(p_1) - \hat{F}_2^{-1}(p_1)\\[1em]
            \vdots \\[1em]
            \hat{F}_1^{-1}(p_J) - \hat{F}_2^{-1}(p_J)
          \end{bmatrix} \xrightarrow{d} \mathcal{N}(0, \Psi) \text{ as } n \to \infty,
    \end{equation*}
    where $\Psi = \Upsilon_{F_1} + \Upsilon_{F_2}$. 
    
    The test statistic, expressed as $\mathfrak{T}_n = \mathcal{Z}_n^T \Psi^{-1} \mathcal{Z}_n$, converges in distribution towards a $\chi_J^2$ as $n \to \infty$ under $\mathcal{H}_0$.
\end{result}
\begin{proof}
    We have under $\mathcal H_0$,
            \begin{equation}
    \mathcal{Z}_n  = 
            \sqrt{n} \begin{bmatrix}
            \hat{F}_1^{-1}(p_1) - F_1^{-1}(p_1) \\[1em]
            \vdots \\[1em]
            \hat{F}_1^{-1}(p_J) - F_1^{-1}(p_J)
        \end{bmatrix} -     \sqrt{n}   \begin{bmatrix}
            \hat{F}_2^{-1}(p_1) - F_2^{-1}(p_1) \\[1em]
            \vdots \\[1em]
            \hat{F}_2^{-1}(p_J) - F_2^{-1}(p_J)
        \end{bmatrix}
    \end{equation}
   Since the two treatment groups are independent, we have from Result 4 that, as $n \to \infty$, $\mathcal{Z}_n \xrightarrow{d} \mathcal{N}\left(0, \Psi \right)$, where $\Psi = \Upsilon_{F_1} + \Upsilon_{F_2}.$ 
    From these results we conclude that under $\mathcal{H}_0$, $\mathfrak{T}_n \xrightarrow{d} \chi_J^2$ as $n \to \infty$.

\end{proof}

\begin{result}
    Under $\mathcal{H}_1: F_1^{-1}(p_j) - F_2^{-1}(p_j) = \Delta_j, j = 1,..., J$, $\mathfrak{T}_n$ is asymptotically equivalent to $\chi_J^2(\Psi^{-1/2} \xi)$, an uncentered chi-squared distribution with $J$ degrees of freedom and mean $\Psi^{-1/2} \xi$, with 
    \begin{equation}
      \xi = \sqrt{n}\begin{bmatrix}
        \Delta_1\\[1em]
                \vdots \\[1em]
         \Delta_J      
     \end{bmatrix}\cdot
    \end{equation}
    We then have the following asymptotic formula for the power of the test of level $\alpha$:
     \begin{equation}
        1 - \beta = F_{\chi_J^2(\Psi^{-1/2} \xi)} (q_{J, 1-\alpha}),
        \label{stat_test_multi}
    \end{equation}
    denoting as $F_{\chi_J^2(\Psi^{-1/2} \xi)}$ the cumulative distribution function of this chi-squared distribution and  $q_{J, 1-\alpha}$ as the quantile of order $1-\alpha$ of the chi-squared distribution with $J$ degrees of freedom.
\end{result}
\begin{proof}
We denote:
\begin{equation}
    Y_n = \sqrt{n}\begin{bmatrix}
        \hat{F}_1^{-1}(p_1) - F_1^{-1}(p_1)\\[1em]
               \vdots \\[1em]
        \hat{F}_1^{-1}(p_J) - F_1^{-1}(p_J)       
    \end{bmatrix} - \sqrt{n}\begin{bmatrix}
        \hat{F}_2^{-1}(p_1) - F_2^{-1}(p_1)\\[1em]
               \vdots \\[1em]
        \hat{F}_2^{-1}(p_J) - F_2^{-1}(p_J)       
    \end{bmatrix} \cdot
\end{equation}
    Under $\mathcal{H}_1$, we write:
\begin{equation}
    \mathcal{Z}_n = Y_n + \xi.
\end{equation}
From Lemma 1 in Kosorok~\cite{kosorok1999two}, we have:
\begin{equation}
    Y_n = \sqrt{n} \begin{bmatrix}
           \dfrac{\hat{F}_1(F_1^{-1}(p_1)) - p_1}{f_1(F_1^{-1}(p_1))}\\[1em]
           \vdots \\[1em]
           \dfrac{\hat{F}_1(F_1^{-1}(p_J)) - p_J}{f_1(F_1^{-1}(p_J))}
         \end{bmatrix} -     \sqrt{n} \begin{bmatrix}
           \dfrac{\hat{F}_2(F_2^{-1}(p_1)) - p_1}{f_2(F_2^{-1}(p_1))}\\[1em]
           \vdots \\[1em]
           \dfrac{\hat{F}_2(F_2^{-1}(p_J)) - p_J}{f_2(F_2^{-1}(p_J))}
         \end{bmatrix} + o_p(1/\sqrt{n}).
\end{equation}
The first and second term on the right-hand side of this equation converge in distribution respectively to $\mathcal{N}(0, \Upsilon_{F_1}) $  and $\mathcal{N}(0, \Upsilon_{F_2})$ as $n \to \infty$. Therefore, 
\begin{equation}
    Y_n \xrightarrow{d}\mathcal{N}(0,\Psi ), \text{ as } n \to \infty,
\end{equation}
and $\mathcal{Z}_n$ is asymptotically equivalent to a $\mathcal{N}(\xi, \Psi)$. 
From the consistency of $\hat\Psi$, $\mathcal{Z}_n^T \hat{\Psi}^{-1} \mathcal{Z}_n$ is asymptotically equivalent to $ \chi_J^2(\Psi^{-1/2} \xi)$. We conclude from the definition of the power,
    \begin{equation}
        \begin{aligned}
            1 - \beta &= P_{\mathcal{H}_1}(\mathfrak{T}_n > q_{J, 1-\alpha}) \\
            &\approx F_{\chi_J^2(\Psi^{-1/2} \xi)} (q_{J, 1-\alpha}).
        \end{aligned}
    \end{equation}

\end{proof}

\section{Details on the estimation of the density}

In the main document, two different methods are proposed for the estimation of $f_k(F_k^{-1}(p))$, $k=1,2$: a resampling procedure based on the method from Lin \textit{et al.}~\cite{lin2015conditional} and a kernel density estimator. The first method only estimates the densities at one point, the quantile $F_k^{-1}(p)$, while the second method estimates the whole function $f_k(t)$ from which the estimator at the quantile is obtained by setting $t=F_k^{-1}(p)$. In Section~\ref{sec:resample} the resampling procedure is explained, while details for the kernel density estimator are provided in Section~\ref{sec:kde}. 

\subsection{Resampling procedure}\label{sec:resample}
We propose a resampling procedure that allows to estimate $f(F^{-1}(p))$ the density at a given quantile, for a given probability $0 < p < 1$.
This method, inspired by Lin \textit{et al.}~\cite{lin2015conditional} and presented in Farah \textit{et al.}~\cite{farah2025note}, consists of generating multiple samples of the zero-mean Gaussian variable from which a least square estimator is constructed.
We require the densities at the quantiles to be strictly positive, and we denote as $\hat{F}$ the consistent estimator for $F$ obtained from the usual Kaplan-Meier estimation. Taking this estimator we obtain $\hat{F}^{-1}(p)$ the estimators of the inverse distribution at $p$. Then we propose the following resampling procedure:

\begin{enumerate}
    \item Generate B realizations of the Gaussian $\varepsilon \sim \mathcal{N}(0, \sigma_{\varepsilon}^2)$, denoted by $\varepsilon_1,..., \varepsilon_B$
    \item Calculate $\sqrt{n}\left( \hat{F}\left(\hat{F}^{-1}(p) + \dfrac{\varepsilon_b}{\sqrt{n}}\right) - p \right), b = 1,..., B$ and denote them as $y_b$, then the least squares estimate of $f(F^{-1}(p))$ is $\hat{A} = (x'x)^{-1}x'Y$, where $x= (\varepsilon_1,..., \varepsilon_B)^T$ and $Y = (y_1,..., y_B)^T$
    
\end{enumerate}

We advocate that the variance of the Gaussian variables must be carefully chosen as it may impact the quality of estimation of the density. 
A grid-search algorithm is used for automatic variance selection in practical applications.
Details about the resampling-based procedure, variance selection and theoretical results on the consistency of the estimator are provided in Farah \textit{et al.}~\cite{farah2025note}.

\FloatBarrier
\subsection{Kernel density estimation}\label{sec:kde}

In the presence of censoring, a typical kernel estimator of the density can be constructed by estimating the censoring distribution from the Kaplan-Meier estimator. In F{\"o}ldes {\it et al.}~\cite{foldes1981strong}, Diehl and Stute~\cite{diehl1988kernel} the following estimator has been proposed:
\begin{align*}
\hat{f}_h(t) = \frac{1}{nh} \sum_{i=1}^n \frac{\delta_i}{\hat S_{\text{cens}}(T_i)} K\left( \frac{T_i - t}{h} \right),
\end{align*}
where $\hat S_{\text{cens}}$ is the Kaplan-Meier estimator of the censoring survival function and $K$ a kernel satisfying standard conditions. In order to compute this estimator, the kernel and the bandwidth must be chosen. In our implementation, a Gaussian kernel was taken and the bandwidth was obtained from cross-validation. For the choice of the bandwidth, the goal of the method is to try to minimize the Integrated Squared Error (ISE), defined, for $k=1, 2$, as: 
\begin{align*}
\text{ISE}(\hat f_h) & = \int \left(\hat f_h(t)-f_k(t)\right)^2dt\\
 & = \int \hat f^2_h(t)dt-2 \int \hat f_h(t)f_k(t)dt  + \int f_k^2(t)dt.
\end{align*}
In this expression the last term does not depend on $h$ and can be omitted. On the other hand, the term $\int f_h(t)f_k(t)dt$ is estimated by:
\begin{align*}
\hat J(h) = \frac{1}{n(n-1)h}\sum_{i\neq j}K\left(\frac{T_i-T_j}{h}\right)\frac{\delta_{ik}\delta_{jk}}{\hat S_{\text{cens}}(T_i)\hat S_{\text{cens}}(T_j)}\cdot
\end{align*}
Using the consistency of the Kaplan-Meier estimator, it can easily be shown that this estimator converges in probability, as $n$ tends to infinity, towards $\int \hat f_h(t)f_k(t)dt$ (see Marron and Padgett~\cite{marron1987asymptotically}). In conclusion, our cross-validated estimator is defined as the Gaussian kernel estimator with bandwidth chosen as the minimizer of $\int \hat f^2_h(t)dt-2 \hat J(h)$, that is:
\begin{align*}
\hat h = \argmin_h \left\{ \int \hat f^2_h(t)dt- \frac{2}{n(n-1)h}\sum_{i\neq j}K\left(\frac{T_i-T_j}{h}\right)\frac{\delta_{ik}\delta_{jk}}{\hat S_{\text{cens}}(T_i)\hat S_{\text{cens}}(T_j)}\right\}\cdot
\end{align*}
We refer the reader to Marron and Padgett~\cite{marron1987asymptotically} for more details about the cross-validation bandwidth selector and theoretical results regarding its validity.

    \section{Additional simulations}
    In this section, we present additional simulations for the Section~3.1 of the main document, in the context of planning clinical trials.
    We report the analytical power and type I error, as well as the empirical rejection rate obtained through ten thousand simulations.
    We recall the simulation scenarios, in which the survival time in the control arm follows an exponential distribution with rate $\lambda_a$, and in the experimental arm it is specified as follows:
    \begin{itemize}
    \item \underline{ Scenario 1 (proportional hazards):} Exponential with rate $\lambda_b$. 
    \item \underline{ Scenario 2 (late differences):} Piecewise exponential with rate $\lambda_a$ until time $t_{\text{cut}}$ and $\lambda_b$ onward.
\end{itemize}
In all scenarios, the distribution of censoring time is exponential with rate $\lambda_{\text{cens}}$. Using the expression derived in the previous section, it is possible to compute the analytical power as a function of the parameters for each scenario. 

Mean computational time per iteration is reported in seconds.
All simulations were conducted on a workstation with an Intel Core i5-13600H CPU and 32 GB of RAM.
Computation times were measured using $\texttt{system.time}$ from $\texttt{R}$ and reported as the elapsed wall-clock time.
The code required to reproduce the results reported here was implemented in R and is publicly available at \url{https://github.com/beafarah/dens-estimation-at-quantile/}.

\newpage    
\subsection{Test at $p=0.5$, 25\% of censoring}
\label{sim_05_25cens}
    In this scenario, we set $\lambda_a = 1.5$, $\lambda_{\text{cens}} = 0.48$, $t_{\text{cut}} = 0.2$.
    \begin{table}[h]
\small\sf\centering
\begin{tabular}{lccc|ccc}
\toprule
\multirow{2}{*}{\textbf{$\Delta$}} & \multicolumn{3}{c|}{\textbf{Scenario 1}} & \multicolumn{3}{c}{\textbf{Scenario 2}} \\
\cmidrule{2-7}
& \textbf{Empirical} & \textbf{Formula} & \textbf{Comp. time} & \textbf{Empirical} & \textbf{Formula} & \textbf{Comp. time} \\
\midrule
\multicolumn{7}{c}{\textbf{Sample size $n_i = 50$}} \\
\midrule
0   & 0.0103 & 0.05   & $1.3 \times 10^{-2} (8.46 \times 10^{-3})$ & 0.0015  & 0.05   & $9.16 \times 10^{-3} (9.41 \times 10^{-3}) $ \\
0.1 & 0.0504 & 0.1236 & $ 1.01 \times 10^{-2} (8.88 \times 10^{-3})$ & 0.0174  & 0.1153 &  $1.28 \times 10^{-2} (1.54 \times 10^{-2}) $\\  
0.2 & 0.3507 & 0.4147 & $1.19 \times 10^{-2} (9.81 \times 10^{-3}) $ & 0.3908  & 0.4682 & $1.36 \times 10^{-2} (1.52 \times 10^{-2})$ \\
\midrule
\multicolumn{7}{c}{\textbf{Sample size $n_i = 100$}} \\
\midrule
0   & 0.0257 & 0.05   & $ 1.3 \times 10^{-2}(1.01 \times 10^{-2}) $ & 0.0060  & 0.05   & $9.5 \times 10^{-3} (8.45 \times 10^{-3} )$ \\
0.1 & 0.1583 & 0.2000 & $1.25 \times 10^{-2} (1.11 \times 10^{-2}) $ & 0.0991  & 0.1831 & $1.34 \times 10^{-2} (2.08 \times 10^{-2})$ \\
0.2 & 0.7280 & 0.6939 & $1.35 \times 10^{-2} ( 8.45 \times 10^{-3}) $ & 0.8645  & 0.7577 & $ 1.49\times 10^{-2} (2.24 \times 10^{-2})$ \\
\midrule
\multicolumn{7}{c}{\textbf{Sample size $n_i = 500$}} \\
\midrule
0   & 0.047  & 0.05   & $1.34 \times 10^{-2} ( 1.05\times 10^{-2}) $ & 0.047 & 0.05   & $1.26 \times 10^{-2} (1.04 \times 10^{-2})$ \\
0.1 & 0.714  & 0.703  & $ 1.75\times 10^{-2} ( 1.83\times 10^{-2}) $ & 0.782  & 0.766  & $1.45 \times 10^{-2} (1.26 \times 10^{-2}) $ \\
0.2 & 1.000  & 1.000  & $ 1.57\times 10^{-2} (1.48 \times 10^{-2}) $ & 1.000  & 1.000  & $1.78 \times 10^{-2} ( 1.73\times 10^{-2}) $ \\
\bottomrule
\end{tabular}
\caption{Type I error and power of the test of equality of quantiles for sample sizes $n_i=50,100,500$, at $p = 0.5$. Mean computational time and standard error per iteration are shown in seconds.}
\label{T1}
\end{table}

\begin{figure}[ht]
    \centering
\begin{tikzpicture}
\begin{groupplot}[
    group style={group size=4 by 1, horizontal sep=0.8cm},
    width=4.2cm, height=4.2cm,
    xmin=0, xmax=3, ymin=0, ymax=1,
    xlabel={}, ylabel={},
    label style={font=\small},           
    tick label style={font=\tiny},
    title style={font=\small},
]
\nextgroupplot[xlabel={Time},ylabel={Survival}, title={Scenario 1, $\Delta = 0.1$}, ymin = -0.1, ymax = 1.1, ytick={0,0.25,0.5,0.75,1.0}]
\addplot[red, thick, domain=0:3]{exp(-1.5*x)};
\addplot[blue, dashed, thick, domain=0:0.2]{exp(-1.5*x)};
\addplot[blue, dashed, thick, domain=0.2:3]{exp(-1.5*0)*exp(-1.9142*(x))};
\addplot[black, dashed] coordinates {(0,0.5) (3,0.5)};

\nextgroupplot[xlabel={Time},title={Scenario 1, $\Delta = 0.2$},ymin = -0.1, ymax = 1.1, ytick={0,0.25,0.5,0.75,1.0}]
\addplot[red, thick, domain=0:3]{exp(-1.5*x)};
\addplot[blue, dashed, thick, domain=0:0.3]{exp(-1.5*x)};
\addplot[blue, dashed, thick, domain=0.3:3]{exp(-1.5*0)*exp(-2.6446*(x))};
\addplot[black, dashed] coordinates {(0,0.5) (3,0.5)};

\nextgroupplot[xlabel={Time}, title={Scenario 2, $\Delta = 0.1$},ymin = -0.1, ymax = 1.1, ytick={0,0.25,0.5,0.75,1.0}]
\addplot[red, thick, domain=0:3]{exp(-1.5*x)};
\addplot[blue, dashed, thick, domain=0:0.15]{exp(-1.5*x)};
\addplot[blue, dashed, thick, domain=0.15:3]{exp(-1.5*0.2)*exp(-2.425*(x-0.2))};
\addplot[black, dashed] coordinates {(0,0.5) (3,0.5)};

\nextgroupplot[xlabel={Time}, title={Scenario 2, $\Delta = 0.2$}, ymin = -0.1, ymax = 1.1, ytick={0,0.25,0.5,0.75,1.0}]
\addplot[red, thick, domain=0:3]{exp(-1.5*x)};
\addplot[blue, dashed, thick, domain=0:0.25]{exp(-1.5*x)};
\addplot[blue, dashed, thick, domain=0.25:3]{exp(-1.5*0.2)*exp(-6.331*(x-0.2))};
\addplot[black, dashed] coordinates {(0,0.5) (3,0.5)};

\end{groupplot}
\end{tikzpicture}
\caption{True survival curves for the test at $p = 0.5$ with 25\% of censoring.}
    \label{fig:survival_curves}
\end{figure}

\newpage
\FloatBarrier
\subsection{Test at $p = 0.75$, 25\% of censoring}
We illustrate the test at a late quantile, closer to the tail of the distributions. 
We set $\lambda_a = 1.5$, $\lambda_{\text{cens}} = 0.48$, $t_{\text{cut}} = 0.2$.
\begin{table}[h]
\small\sf\centering
\begin{tabular}{lccc|ccc}
\toprule
\multirow{2}{*}{\textbf{$\Delta$}} & \multicolumn{3}{c|}{\textbf{Scenario 1}} & \multicolumn{3}{c}{\textbf{Scenario 2}} \\
\cmidrule{2-7}
& \textbf{Empirical} & \textbf{Formula} & \textbf{Comp. time} & \textbf{Empirical} & \textbf{Formula} & \textbf{Comp. time} \\
\midrule
\multicolumn{7}{c}{\textbf{Sample size $n_i = 50$}} \\
0   & 0.0040 & 0.05   & $9.4 \times 10^{-3} (1.18 \times 10^{-2})$ & 0.0051 & 0.05   & $ 1.31\times 10^{-2} (1.60 \times 10^{-2})$ \\
0.1 & 0.0086 & 0.0685 & $ 1.15 \times 10^{-2} ( 1.36\times 10^{-2})$ & 0.0092 & 0.0669 & $ 1.56 \times 10^{-2} (2.23 \times 10^{-2})$ \\
0.2 & 0.0361 & 0.1356 & $ 1.44 \times 10^{-2} (2.07 \times 10^{-2})$ & 0.0306 & 0.1314 &  $ 1.43 \times 10^{-2} (1.55 \times 10^{-2})$ \\
\midrule
\multicolumn{7}{c}{\textbf{Sample size $n_i = 100$}} \\
0   & 0.0109 & 0.05   & $1.3 \times 10^{-2} ( 2.01 \times 10^{ -2})$ & 0.0108 & 0.05   & $ 1.44\times 10^{-2} (1.62 \times 10^{-2})$ \\
0.1 & 0.0330 & 0.0874 & $ 1.42 \times 10^{-2} (1.92 \times 10^{-2})$ & 0.0312 & 0.0841 & $ 1.44 \times 10^{-2} (1.45 \times 10^{-2})$ \\
0.2 & 0.1440 & 0.2243 & $ 1.54 \times 10^{-2} (1.99 \times 10^{-2})$ & 0.1365 & 0.2158 & $ 1.51 \times 10^{-2} (1.69 \times 10^{-2})$ \\
\midrule
\multicolumn{7}{c}{\textbf{Sample size $n_i = 500$}} \\
0   & 0.0355 & 0.05   & $1.82  \times 10^{-2} ( 1.64\times 10^{ -3})$ & 0.0336 & 0.05   & $ 1.82 \times 10^{-2} (2.41 \times 10^{-2})$ \\
0.1 & 0.2288 & 0.2444 & $ 1.68 \times 10^{-2} (2.13 \times 10^{-2})$ & 0.2335 & 0.2270 & $ 1.63 \times 10^{-2} (1.58 \times 10^{-2})$ \\
0.2 & 0.7779 & 0.7650 & $ 1.88 \times 10^{-2} (2.04 \times 10^{-2})$ & 0.8055 & 0.7447 & $ 1.81 \times 10^{-2} (2.02 \times 10^{-2})$ \\
\bottomrule
\end{tabular}
\caption{Type I error and power of the test of equality of quantiles for sample sizes $n_i=50,100,500$, at $p = 0.75$.}
\label{T2}
\end{table}

\begin{figure}[ht]
    \centering
\begin{tikzpicture}
\begin{groupplot}[
    group style={group size=4 by 1, horizontal sep=0.8cm},
    width=4.2cm, height=4.2cm,
    xmin=0, xmax=3, ymin=0, ymax=1,
    xlabel={}, ylabel={},
    label style={font=\small},           
    tick label style={font=\tiny},
    title style={font=\small},
]
\nextgroupplot[title={Scenario 1, $\Delta = 0.1$}, xlabel= {Time}, ylabel = {Survival}, ymin = -0.1, ymax = 1.1, ytick={0,0.25,0.5,0.75,1.0}]
\addplot[red, thick, domain=0:3]{exp(-1.5*x)};
\addplot[blue, dashed, thick, domain=0:0.2]{exp(-1.5*x)};
\addplot[blue, dashed, thick, domain=0.2:3]{exp(-1.5*0)*exp(-1.6819*(x))};
\addplot[black, dashed] coordinates {(0,0.25) (3,0.25)};

\nextgroupplot[title={Scenario 1, $\Delta = 0.2$}, xlabel = {Time}, ymin = -0.1, ymax = 1.1, ytick={0,0.25,0.5,0.75,1.0}]
\addplot[red, thick, domain=0:3]{exp(-1.5*x)};
\addplot[blue, dashed, thick, domain=0:0.3]{exp(-1.5*x)};
\addplot[blue, dashed, thick, domain=0.3:3]{exp(-1.5*0)*exp(-1.914*(x))};
\addplot[black, dashed] coordinates {(0,0.25) (3,0.25)};

\nextgroupplot[title={Scenario 2, $\Delta = 0.1$}, xlabel = {Time}, ymin = -0.1, ymax = 1.1, ytick={0,0.25,0.5,0.75,1.0}]
\addplot[red, thick, domain=0:3]{exp(-1.5*x)};
\addplot[blue, dashed, thick, domain=0:0.15]{exp(-1.5*x)};
\addplot[blue, dashed, thick, domain=0.15:3]{exp(-1.5*0.2)*exp(-1.7403*(x-0.2))};
\addplot[black, dashed] coordinates {(0,0.25) (3,0.25)};

\nextgroupplot[title={Scenario 2, $\Delta = 0.2$}, xlabel = {Time}, ymin = -0.1, ymax = 1.1, ytick={0,0.25,0.5,0.75,1.0}]
\addplot[red, thick, domain=0:3]{exp(-1.5*x)};
\addplot[blue, dashed, thick, domain=0:0.25]{exp(-1.5*x)};
\addplot[blue, dashed, thick, domain=0.25:3]{exp(-1.5*0.2)*exp(-2.072*(x-0.2))};
\addplot[black, dashed] coordinates {(0,0.25) (3,0.25)};

\end{groupplot}
\end{tikzpicture}
\caption{True survival curves for the test at $p = 0.75$ with 25\% of censoring. }
    \label{fig:survival_curves2}
\end{figure}

\newpage
\FloatBarrier
\subsection{Test at $p = 0.9$, 5\% of censoring}
We illustrate the test at a high extreme quantile, at the tail of the distributions.
We set $\lambda_a = 3$, $\lambda_{\text{cens}} = 0.1$, $t_{\text{cut}} = 0.2$.
\begin{table}[h]
\small\sf\centering
{
\begin{tabular}{lccc|ccc}
\toprule
\multirow{2}{*}{\textbf{$\Delta$}} & \multicolumn{3}{c|}{\textbf{Scenario 1}} & \multicolumn{3}{c}{\textbf{Scenario 2}} \\
\cmidrule{2-7}
& \textbf{Empirical} & \textbf{Formula} & \textbf{Comp. time} & \textbf{Empirical} & \textbf{Formula} & \textbf{Comp. time} \\
\midrule
\multicolumn{7}{c}{\textbf{Sample size $n_i = 50$}} \\
0   & 0.0097 & 0.05   & $ 1.61 \times 10^{-2} ( 1.39 \times 10^{-2})$ & 0.0056  & 0.05   & $ 1.18\times 10^{-2} ( 1.343\times 10^{-2})$ \\
0.1 & 0.0232 & 0.081 & $  2.05 \times 10^{-2} ( 2.76\times 10^{-2})$ & 0.0147 & 0.0832  & $ 1.12\times 10^{-2} (1.289 \times 10^{-2})$ \\
0.2 & 0.102  & 0.1991 & $ 1.16 \times 10^{-2} ( 2.05\times 10^{-2})$ & 0.0709 & 0.2126 &  $ 1.35 \times 10^{-2} ( 2.06 \times 10^{-2})$ \\
\midrule
\multicolumn{7}{c}{\textbf{Sample size $n_i = 100$}} \\
0   & 0.0109 & 0.05   & $ 2.48 \times 10^{-2} (2.76  \times 10^{ -2})$ & 0.0103  & 0.05   & $ 1.43\times 10^{-2} ( 1.26\times 10^{-2})$ \\
0.1 & 0.0511 & 0.1142 & $1.47 \times 10^{-2} ( 1.83\times 10^{-2})$ & 0.0469       & 0.1172 & $ 1.07 \times 10^{-2} (1.35 \times 10^{-2})$ \\
0.2 & 0.2559 & 0.349 &  $1.35 \times 10^{-2} (1.45 \times 10^{-2})$ & 0.2555      & 0.3744 & $ 1.92 \times 10^{-2} ( 2.25\times 10^{-2})$ \\
\midrule
\multicolumn{7}{c}{\textbf{Sample size $n_i = 500$}} \\
0   & 0.0339 & 0.05   & $ 1.51\times 10^{-2} ( 1.47\times 10^{-2})$ & 0.0329 & 0.05   & $ 1.56 \times 10^{-2} ( 1.335\times 10^{-2})$ \\
0.1 & 0.3628  & 0.3774 & $ 1.55\times 10^{-2} ( 1.43\times 10^{-2})$ &  0.377    & 0.3916 & $  1.59 \times 10^{-2} ( 1.42 \times 10^{-2})$ \\
0.2 & 0.9349 & 0.9398 & $ 1.56\times 10^{-2} ( 1.21 \times 10^{-2})$ &   0.9602  &  0.9558 & $ 1.72 \times 10^{-2} ( 2.07\times 10^{-2})$ \\
\bottomrule
\end{tabular}
\caption{Type I error and power of the test of equality of quantiles for sample sizes $n_i=50,100,500$, at $p = 0.9$.}
\label{T2_new}
}
\end{table}

\begin{figure}[ht]
    \centering
\begin{tikzpicture}
\begin{groupplot}[
    group style={group size=4 by 1, horizontal sep=0.8cm},
    width=4.2cm, height=4.2cm,
    xmin=0, xmax=1, ymin=0, ymax=1,
    xlabel={}, ylabel={},
    label style={font=\small},           
    tick label style={font=\tiny},
    title style={font=\small},
]
\nextgroupplot[title={Scenario 1, $\Delta = 0.1$}, xlabel = {Time}, ylabel = {Survival}, ymin = -0.1, ymax = 1.1, ytick={0,0.25,0.5,0.75,1.0}]
\addplot[red, thick, domain=0:3]{exp(-3*x)};
\addplot[blue, dashed, thick, domain=0:0.2]{exp(-3*x)};
\addplot[blue, dashed, thick, domain=0.2:3]{exp(-3*0)*exp(-3.449*(x))};
\addplot[black, dashed] coordinates {(0,0.1) (3,0.1)};

\nextgroupplot[title={Scenario 1, $\Delta = 0.2$}, xlabel = {Time}, ymin = -0.1, ymax = 1.1, ytick={0,0.25,0.5,0.75,1.0}]
\addplot[red, thick, domain=0:3]{exp(-3*x)};
\addplot[blue, dashed, thick, domain=0:0.3]{exp(-3*x)};
\addplot[blue, dashed, thick, domain=0.3:3]{exp(-3*0)*exp(-4.057*(x))};
\addplot[black, dashed] coordinates {(0,0.1) (3,0.1)};

\nextgroupplot[title={Scenario 2, $\Delta = 0.1$}, xlabel = {Time}, ymin = -0.1, ymax = 1.1, ytick={0,0.25,0.5,0.75,1.0}]
\addplot[red, thick, domain=0:3]{exp(-3*x)};
\addplot[blue, dashed, thick, domain=0:0.2]{exp(-3*x)};
\addplot[blue, dashed, thick, domain=0.2:3]{exp(-3*0.2)*exp(-3.641*(x-0.2))};
\addplot[black, dashed] coordinates {(0,0.1) (3,0.1)};

\nextgroupplot[title={Scenario 2, $\Delta = 0.2$}, xlabel = {Time}, ymin = -0.1, ymax = 1.1, ytick={0,0.25,0.5,0.75,1.0}]
\addplot[red, thick, domain=0:3]{exp(-3*x)};
\addplot[blue, dashed, thick, domain=0:0.2]{exp(-3*x)};
\addplot[blue, dashed, thick, domain=0.2:3]{exp(-3*0.2)*exp(-4.632*(x-0.2))};
\addplot[black, dashed] coordinates {(0,0.1) (3,0.1)};

\end{groupplot}
\end{tikzpicture}
\caption{True survival curves for the test at $p = 0.05$ with 5\% of censoring.}
    \label{fig:survival_curves3}
\end{figure}

\color{black}

\newpage
\FloatBarrier
\subsection{Test at $p = 0.25$, 25\% of censoring}
We illustrate the test at a low quantile.
We set $\lambda_a = 0.1$, $\lambda_{\text{cens}} = 0.03$, $t_{\text{cut}} = 0.2$.
\begin{table}[h]
\small\sf\centering
\begin{tabular}{lccc|ccc}
\toprule
\multirow{2}{*}{\textbf{$\Delta$}} & \multicolumn{3}{c|}{\textbf{Scenario 1}} & \multicolumn{3}{c}{\textbf{Scenario 2}} \\
\cmidrule{2-7}
& \textbf{Empirical} & \textbf{Formula} & \textbf{Comp. time} & \textbf{Empirical} & \textbf{Formula} & \textbf{Comp. time} \\
\midrule
\multicolumn{7}{c}{\textbf{Sample size $n_i = 50$}} \\
0   & 0.0145 & 0.05   & $8.84 \times 10^{-3} (9.96 \times 10^{-3}) $  & 0.0198 & 0.05   &  $ 1.04\times 10^{-2} (9.62 \times 10^{-3}) $ \\
0.1 & 0.0173 & 0.0502 & $ 9.94 \times 10^{-3} (9.63 \times 10^{-3}) $ & 0.0190 & 0.0504 &  $ 1.14\times 10^{-2} ( 1.11\times 10^{-2}) $ \\
0.2 & 0.0158 & 0.0508 & $ 8.46 \times 10^{-3} ( 8.52\times 10^{-3}) $ & 0.0188 & 0.0171 &  $ 1.38 \times 10^{-2} ( 1.01\times 10^{-2}) $ \\
\midrule
\multicolumn{7}{c}{\textbf{Sample size $n_i = 100$}} \\
0   & 0.0304 & 0.05   & $ 9.5 \times 10^{-3} (9.25 \times 10^{-3}) $ & 0.0331 & 0.05   &  $1.38 \times 10^{-2} (1.00 \times 10^{-2}) $ \\
0.1 & 0.0287 & 0.0504 & $ 9.4 \times 10^{-3} (8.74 \times 10^{-3}) $ & 0.0383 & 0.0508 &  $ 1.14\times 10^{-2} (1.16 \times 10^{-2}) $ \\
0.2 & 0.0349 & 0.0516 & $ 1.06 \times 10^{-2} (9.30 \times 10^{-3}) $ & 0.0377 & 0.0534 &  $ 1.27\times 10^{-2} (9.65 \times 10^{-3}) $ \\
\midrule
\multicolumn{7}{c}{\textbf{Sample size $n_i = 500$}} \\
0   & 0.0466 & 0.05   & $1.17 \times 10^{-2} (1.22 \times 10^{-2}) $ & 0.0475 & 0.05   &  $ 1.40 \times 10^{-2} ( 1.12\times 10^{-2}) $ \\
0.1 & 0.0484 & 0.0520 & $1.28 \times 10^{-2} ( 7.92\times 10^{-3}) $ & 0.0550 & 0.0540 &  $ 1.57\times 10^{-2} ( 9.97\times 10^{-3}) $ \\
0.2 & 0.0526 & 0.0582 & $1.41 \times 10^{-2} (7.93 \times 10^{-3}) $ & 0.0780 & 0.0673 &  $ 1.16\times 10^{-2} (9.38 \times 10^{-3}) $ \\
\bottomrule
\end{tabular}
\caption{Type I error and power of the test of equality of quantiles for sample sizes $n_i=50,100,500$, at $p = 0.25$. }
\label{T3}
\end{table}

\begin{figure}[ht]
    \centering
\begin{tikzpicture}
\begin{groupplot}[
    group style={group size=4 by 1, horizontal sep=0.8cm},
    width=4.2cm, height=4.2cm,
    xmin=0, xmax=5, ymin=0.55, ymax=1.05,
    xlabel={}, ylabel={},
    label style={font=\small},           
    tick label style={font=\tiny},
    title style={font=\small},
]
\nextgroupplot[title={Scenario 1, $\Delta = 0.1$}, xlabel = {Time}, ylabel = {Survival}]
\addplot[red, thick, domain=0:5]{exp(-0.1*x)};
\addplot[blue, dashed, thick, domain=0:0.2]{exp(-0.1*x)};
\addplot[blue, dashed, thick, domain=0.2:5]{exp(-0.1*0)*exp(-0.1036*(x))};
\addplot[black, dashed] coordinates {(0,0.75) (5,0.75)};

\nextgroupplot[title={Scenario 1, $\Delta = 0.2$}, xlabel = {Time}]
\addplot[red, thick, domain=0:5]{exp(-0.1*x)};
\addplot[blue, dashed, thick, domain=0:0.2]{exp(-0.1*x)};
\addplot[blue, dashed, thick, domain=0.2:5]{exp(-0.1*0)*exp(-0.1074*(x))};
\addplot[black, dashed] coordinates {(0,0.75) (5,0.75)};

\nextgroupplot[title={Scenario 2, $\Delta = 0.1$}, xlabel = {Time}]
\addplot[red, thick, domain=0:5]{exp(-0.1*x)};
\addplot[blue, dashed, thick, domain=0:0.2]{exp(-0.1*x)};
\addplot[blue, dashed, thick, domain=0.2:5]{exp(-0.1*0.2)*exp(-0.1038*(x-0.2))};
\addplot[black, dashed] coordinates {(0,0.75) (5,0.75)};

\nextgroupplot[title={Scenario 2, $\Delta = 0.2$}, xlabel = {Time}]
\addplot[red, thick, domain=0:5]{exp(-0.1*x)};
\addplot[blue, dashed, thick, domain=0:0.2]{exp(-0.1*x)};
\addplot[blue, dashed, thick, domain=0.2:5]{exp(-0.1*0.2)*exp(-0.1080*(x-0.2))};
\addplot[black, dashed] coordinates {(0,0.75) (5,0.75)};

\end{groupplot}
\end{tikzpicture}
\caption{True survival curves for the test at $p = 0.05$ with 25\% of censoring.}
    \label{fig:survival_curvesb}
\end{figure}

\newpage
\FloatBarrier
\subsection{Test at $p = 0.05$, 25\% of censoring}
We show simulations for a low extreme quantile of survival. In order to do so, we let $\lambda_a = 0.1$, $\lambda_{\text{cens}} = 0.03$, $t_{\text{cut}} = 0.2$.
\begin{table}[h]
\small\sf\centering
{
\begin{tabular}{lccc|ccc}
\toprule
\multirow{2}{*}{\textbf{$\Delta$}} & \multicolumn{3}{c|}{\textbf{Scenario 1}} & \multicolumn{3}{c}{\textbf{Scenario 2}} \\
\cmidrule{2-7}
& \textbf{Empirical} & \textbf{Formula} & \textbf{Comp. time} & \textbf{Empirical} & \textbf{Formula} & \textbf{Comp. time} \\
\midrule
\multicolumn{7}{c}{\textbf{Sample size $n_i = 50$}} \\
0   & 0.0122 & 0.05   & $ 1.55\times 10^{-2} ( 1.149\times 10^{-2}) $  & 0.0117 & 0.05   &  $ 1.24\times 10^{-2} (1.31\times 10^{-2}) $ \\
0.1 & 0.0156 & 0.0565 & $ 1.5\times 10^{-2} ( 1.487\times 10^{-2}) $         &   0.0153    &   0.0512     &  $ 1.01 \times 10^{-2} (1.184 \times 10^{-2}) $ \\
0.2 & 0.0201 & 0.0820 & $  1.59\times 10^{-2} (1.949\times 10^{-2}) $         &   0.0240    &  0.0627       &  $ 1.41  \times 10^{-2} ( 1.752\times 10^{-2}) $ \\
\midrule
\multicolumn{7}{c}{\textbf{Sample size $n_i = 100$}} \\
0   & 0.0281 & 0.05   & $ 1.55 \times 10^{-2} (1.076 \times 10^{-2}) $ & 0.0254 & 0.05   &  $ 1.3\times 10^{-2} (1.84 \times 10^{-2}) $ \\
0.1 & 0.0356 & 0.0632     & $  1.58 \times 10^{-2} ( 1.199\times 10^{-2}) $ & 0.0281 & 0.0524 &  $ 1.04\times 10^{-2} ( 1.994\times 10^{-2}) $ \\
0.2 &  0.0613  & 0.11499       & $1.37 \times 10^{-2} (1.169 \times 10^{-2}) $    &  0.0517        &  0.0757 &  $ 1.2\times 10^{-2} ( 1.39\times 10^{-2}) $ \\
\midrule
\multicolumn{7}{c}{\textbf{Sample size $n_i = 500$}} \\
0   & 0.0491 & 0.05   & $ 1.7\times 10^{-2} ( 1.82\times 10^{-2})$  & 0.049 & 0.05   &  $  1.46 \times 10^{-2} ( 1.36\times 10^{-2}) $ \\
0.1 & 0.1073 & 0.1176 & $ 1.4\times 10^{-2} ( 1.23\times 10^{-2})$  &     0.1103   & 0.1264 &  $ 1.36 \times 10^{-2} (1.466\times 10^{-2}) $ \\
0.2 & 0.3771 & 0.3816  & $ 1.37 \times 10^{-2} ( 1.3 \times 10^{-2})$  &     0.423   & 0.4470 &  $ 1.23\times 10^{-2} ( 1.118\times 10^{-2}) $ \\
\bottomrule
\end{tabular}
}
\caption{Type I error and power of the test of equality of quantiles for sample sizes $n_i=50,100,500$, at $p = 0.05$. }
\label{T3_new}
\end{table}

\begin{figure}[ht]
    \centering
\begin{tikzpicture}
\begin{groupplot}[
    group style={group size=4 by 1, horizontal sep=0.8cm},
    width=4.2cm, height=4.2cm,
    xmin=0, xmax=1.5, ymin=0.75, ymax=1.05,
    xlabel={}, ylabel={},
    label style={font=\small},           
    tick label style={font=\tiny},
    title style={font=\small},
]
\nextgroupplot[title={Scenario 1, $\Delta = 0.1$}, xlabel = {Time}, ylabel = {Survival}]
\addplot[red, thick, domain=0:1.5]{exp(-0.1*x)};
\addplot[blue, dashed, thick, domain=0:0.2]{exp(-0.1*x)};
\addplot[blue, dashed, thick, domain=0.2:1.5]{exp(-0.1*0)*exp(-0.1242*(x))};
\addplot[black, dashed] coordinates {(0,0.95) (1.5,0.95)};

\nextgroupplot[title={Scenario 1, $\Delta = 0.2$}, xlabel = {Time}]
\addplot[red, thick, domain=0:1.5]{exp(-0.1*x)};
\addplot[blue, dashed, thick, domain=0:0.2]{exp(-0.1*x)};
\addplot[blue, dashed, thick, domain=0.2:1.5]{exp(-0.1*0)*exp(-0.1639*(x))};
\addplot[black, dashed] coordinates {(0,0.95) (1.5,0.95)};

\nextgroupplot[title={Scenario 2, $\Delta = 0.1$}, xlabel = {Time}]
\addplot[red, thick, domain=0:1.5]{exp(-0.1*x)};
\addplot[blue, dashed, thick, domain=0:0.2]{exp(-0.1*x)};
\addplot[blue, dashed, thick, domain=0.2:1.5]{exp(-0.1*0.2)*exp(-0.1469*(x-0.2))};
\addplot[black, dashed] coordinates {(0,0.95) (1.5,0.95)};

\nextgroupplot[title={Scenario 2, $\Delta = 0.2$}, xlabel = {Time}]
\addplot[red, thick, domain=0:1.5]{exp(-0.1*x)};
\addplot[blue, dashed, thick, domain=0:0.2]{exp(-0.1*x)};
\addplot[blue, dashed, thick, domain=0.2:1.5]{exp(-0.1*0.2)*exp(-0.2770*(x-0.2))};
\addplot[black, dashed] coordinates {(0,0.95) (1.5,0.95)};

\end{groupplot}
\end{tikzpicture}
\caption{True survival curves. First and second columns correspond to Scenario 1, third and fourth columns correspond to Scenario 2. }
    \label{fig:survival_curvesc}
\end{figure}
\color{black}

\newpage
\FloatBarrier
\subsection{Test at $p=0.5$, 50\% of censoring}
We modify scenario \ref{sim_05_25cens}, now with a higher censoring rate.
  \begin{table}[h]
\small\sf\centering
\begin{tabular}{lccc|ccc}
\toprule
\multirow{2}{*}{\textbf{$\Delta$}} & \multicolumn{3}{c|}{\textbf{Scenario 1}} & \multicolumn{3}{c}{\textbf{Scenario 2}} \\
\cmidrule{2-7}
& \textbf{Empirical} & \textbf{Formula} & \textbf{Comp. time} & \textbf{Empirical} & \textbf{Formula} & \textbf{Comp. time} \\
\midrule
\multicolumn{7}{c}{\textbf{Sample size $n_i = 50$}} \\
0   & 0.0043 & 0.05   & $ 1.11 \times 10^{-2} (1.25 \times 10^{-2}) $  & 0.0044 & 0.05   & $8.54 \times 10^{-3} ( 8.79\times 10^{-3}) $ \\
0.1 & 0.0229 & 0.1112 & $ 1.17\times 10^{-2} (1.15 \times 10^{-2}) $ & 0.0084 & 0.1071 & $9.04 \times 10^{-3} ( 9.86\times 10^{-3}) $ \\
0.2 & 0.2251 & 0.3586 & $ 9.48\times 10^{-3} (9.60 \times 10^{-3}) $ & 0.1650 & 0.4032 & $ 9.22\times 10^{-3} (1.06 \times 10^{-2}) $ \\
\midrule
\multicolumn{7}{c}{\textbf{Sample size $n_i = 100$}} \\
0   & 0.0149 & 0.05   & $ 1.18\times 10^{-2} ( 9.64\times 10^{-3}) $ & 0.0167 & 0.05   & $ 9.38\times 10^{-3} (9.40 \times 10^{-3}) $ \\
0.1 & 0.1102 & 0.1748 & $1.16 \times 10^{-2} ( 1.07\times 10^{-2}) $ & 0.1021 & 0.1663 & $8.72 \times 10^{-3} ( 9.28\times 10^{-3}) $ \\
0.2 & 0.6461 & 0.6178 & $ 1.03\times 10^{-2} (9.63 \times 10^{-3}) $ & 0.8340 & 0.6790 & $ 1.04\times 10^{-2} ( 1.05\times 10^{-2}) $ \\
\midrule
\multicolumn{7}{c}{\textbf{Sample size $n_i = 500$}} \\
0   & 0.0470 & 0.05   & $ 1.30\times 10^{-2} ( 9.13\times 10^{-3}) $ & 0.0436 & 0.05   & $ 1.34\times 10^{-2} ( 9.89 \times 10^{-3}) $ \\
0.1 & 0.6296 & 0.6249 & $1.24 \times 10^{-2} (9.27 \times 10^{-3}) $ & 0.7010 & 0.5954 & $ 1.36\times 10^{-2} (9.89 \times 10^{-3}) $ \\
0.2 & 1.0000 & 1.0000 & $ 1.46\times 10^{-2} (1.00 \times 10^{-2}) $ & 1.0000 & 1.0000 & $ 1.30\times 10^{-2} (9.94 \times 10^{-3}) $ \\
\bottomrule
\end{tabular}
\caption{Type I error and power of the test of equality of quantiles for sample sizes $n_i=50,100,500$, at $p = 0.5$, with more censoring. }
\label{T4}
\end{table}
